\documentclass[journal]{mock-aiaa}
\usepackage[utf8]{inputenc}

\usepackage{graphicx}
\usepackage{amsmath,bm,mathtools}
\usepackage{amssymb,color,array,natbib,url}
\usepackage{siunitx}
\usepackage{longtable,tabularx}
\title{NASA's Meteoroid Engineering Model (MEM) 3 \\ and its ability to replicate spacecraft impact rates}

\author{Althea V.~Moorhead}
\affil{NASA Meteoroid Environment Office, Marshall Space Flight Center EV44, Alabama, 35812}
\author{Aaron Kingery}
\affil{ERC, Jacobs Space Exploration Group, NASA Meteoroid Environment Office \\ Marshall Space Flight Center, Huntsville, AL 35812 USA}
\author{Steven Ehlert}
\affil{Qualis Corporation, Jacobs Space Exploration Group, NASA Meteoroid Environment Office \\ Marshall Space Flight Center, Huntsville, AL 35812 USA}

\begin{document}

\twocolumn[
  \begin{@twocolumnfalse}
    \maketitle

\begin{abstract}
Meteoroids pose one of the largest risks to spacecraft outside of low Earth orbit. In order to correctly predict the rate at which meteoroids impact and damage spacecraft,  environment models must describe the mass, directionality, velocity, and density distributions of meteoroids. NASA's Meteoroid Engineering Model (MEM) is one such model; MEM~3 is an updated version of the code that better captures the correlation between directionality and velocity and incorporates a bulk density distribution. This paper describes MEM~3 and compares its predictions with the rate of large particle impacts seen on the Long Duration Exposure Facility (LDEF) and the Pegasus II and III satellites.
\end{abstract}
\vspace{0.333in}

  \end{@twocolumnfalse}
]

\section*{Nomenclature}

%{\renewcommand\arraystretch{1.0}
%\noindent\begin{longtable*}{@{}l @{\quad=\quad} l@{}}
\begin{tabular}{@{}l @{\quad=\quad} p{6.5cm}}
$a$ & semimajor axis \\
BH & Brinell hardness \\
$b$ & unitless parameter that relates $\Delta$, $y$, $x$, and $t_t$ \\
$c$ & speed of sound in meteoroid \\
$c_{0,t}$ & speed of sound in unstressed target material \\
$c_t$ & speed of sound in target \\
$d$ & meteoroid diameter \\
$d_0$ & crater diameter without supralinearity correction \\
$d_c$ & crater diameter \\
$E$ & Young's modulus \\
$E_t$ & Young's modulus of target \\
$e$ & orbital eccentricity \\
$F$ & flux \\
$F_c$ & crater- or damage-limited flux \\
$F_m$ & mass-limited flux \\
$F_G$ & Gr\"{u}n et al.~flux \\
$f$ & supralinearity correction \\
$G$ & gravitational constant \\
$h$ & altitude \\
$h_1$ & altitude of 100 km \\
$h_2$ & altitude of 100,000 km \\
$i$ & orbital inclination \\
\end{tabular}
\begin{tabular}{@{}l @{\quad=\quad} p{6.5cm}}
$M_\odot$ & mass of the Sun \\
$M_\oplus$ & mass of the Earth \\
$N_{c,i}$ & number of craters on side $i$ \\
$m$ & meteoroid mass \\
$P$ & probability \\
$p_c$ & crater depth \\
$Q$ & aphelion distance \\
$q$ & perihelion distance \\
$R_\oplus$ & radius of the Earth \\
$r$ & heliocentric distance \\
$s_t$ & stress factor of target \\
$t_t$ & target thickness \\
$v$ & meteoroid velocity \\
$v_\bot$ & normal velocity \\
$v_0$ & minimum speed required to produce a crater \\
$v_1$ & speed at 100 km \\
$v_2$ & speed at 100,000 km \\
$v_{esc}$ & local escape velocity \\
$v_f$ & meteoroid speed with gravitational focusing \\
$v_i$ & meteoroid speed without gravitational focusing \\
$x$ & ratio of uncorrected crater diameter $d_0$ to meteoroid diameter $d$ \\
$Y_t$ & yield strength of target \\
$y$ & unitless parameter that relates $t_t$, $f$, and $d$ \\
\end{tabular}
\begin{tabular}{@{}l @{\quad=\quad} p{6.5cm}}
$z$ & unitless parameter that relates $y$, $t_t$, and $d$ \\
$\alpha_{i,j}$ & angle between surface normal vector $i$ and meteoroid radiant $j$ \\
$\Delta$ & grain size parameter \\
$\bar{\eta}_g$ & average gravitational focusing factor \\
$\theta$ & azimuthal angle \\
$\mu$ & mean of a normal distribution \\
$\rho$ & meteoroid density \\
$\rho_t$ & target density \\
$\sigma$ & standard deviation of a normal distribution \\
$\sigma_t$ & ultimate strength of target \\
$\psi$ & angle between the velocity vector and the radius vector \\
$\phi$ & elevation angle \\
$\xi$ & depth-to-diameter ratio \\
%\end{longtable*}}
\end{tabular}

\section{Introduction}
\label{sec:intro}

\lettrine{M}{eteoroid} impacts threaten spacecraft and astronauts at all locations within the Solar System. At certain altitudes in low Earth orbit, orbital debris are the primary driver of risk, but meteoroids dominate at altitudes below 250~km and above 4000~km \cite{Cooke2017}. In interplanetary space, orbital debris is nonexistent and meteoroids constitute the entire population of potentially dangerous impactors.

NASA's Meteoroid Environment Office (MEO) created the Meteoroid Engineering Model \cite{2004EM&P...95..123M} to assist spacecraft engineers in assessing the risk posed by meteoroids. MEM models the meteoroid background component, which meteor astronomers term the ``sporadic complex.'' The sporadic component comprises the vast majority of the meteoroid environment at sizes that are potentially threatening to spacecraft (i.e., those between 100-200~$\mu$m and 1~cm); meteor showers contribute somewhere between 1\% and 5\% of the total flux, depending on the limiting size \cite{Moorhead2017}. Unlike meteor showers, which are brief in duration, sporadic meteoroids can impact spacecraft at any time during the year. 

Meteoroid impacts can induce a variety of effects, including cratering and/or spalling, severing wires or antennae, producing attitude disturbances, and generating conductive plasmas that can prompt harmful electrical discharges. Each effect has a different dependence on impactor mass or size, density, impact angle, and relative speed. Therefore, meteoroid environment models such as MEM must fully describe the directionality, velocity, density, and mass distribution of meteoroids.

In this paper, we describe how MEM models the meteoroid environment and the recent improvements the MEO has made to the code in order to better support space missions. The most significant improvements are:
\begin{itemize}
\item a meteoroid bulk density model based on recent measurements \cite{2011A&A...530A.113K,2017MNRAS.472.3833M},
\item the ability to analyze transfer trajectories in a single run,
\item the ability to analyze trajectories near Mars, Mercury, and Venus,
\item the correction of certain errors,
\item a cross-platform command-line executable version as well as a GUI,
\item increased user control over run options, and
\item a roughly three-fold improvement in run time.
\end{itemize}
All significant improvements are discussed in detail in Section \ref{sec:model}.

We have tested MEM~3 against the two best sets of meteoroid impact data in near-Earth space: the Pegasus satellites and the Long Duration Exposure Facility (LDEF). The Pegasus satellites represent the first and largest-scale effort to measure the meteoroid environment and the flux measurements obtained from the Pegasus program have been folded into numerous meteoroid models, including the widely used Gr\"{u}n et al.~model \cite{1985Icar...62..244G}. LDEF dedicated a smaller surface area to measuring meteoroid impacts, but recorded data over a longer period of time (nearly six years for LDEF \cite{1990LPI....21.1385Z} as compared to less than a year for Pegasus \cite{Clifton:1966uw}). Furthermore, LDEF maintained a fixed orientation relative to its orbit for the entirely of its lifetime, providing the possibility of probing the meteoroid directionality and speed distributions \cite{1991NASCP3134..569Z}. We test MEM~3 against both satellites in Section \ref{sec:insitu}.

\section{Model description}
\label{sec:model}

MEM is based on various studies of the near-Earth meteoroid environment and the interplanetary dust population. Its mass distribution is based on fireball observations and \emph{in situ} meteoroid impacts \cite{1985Icar...62..244G}, its directionality and velocity are based on radar observations of small meteors \cite{Jones:2004uw}, its heliocentric distance dependence is based on zodiacal light observations \cite{1983A&A...118..345L}, and meteoroid bulk densities are based on meteoroid ablation modeling \cite{2011A&A...530A.113K,2017MNRAS.472.3833M}. This section describes each component of the model and any corresponding improvements.

\subsection{Mass distribution}
\label{sec:mass}

Gr\"{u}n et al.~\cite{1985Icar...62..244G} developed a model of the mass distribution of meteoroids near 1~au by simulating the physical processes experienced by small particles and calibrating the result against observations. Collisions tend to grind down particles and radiative forces modify their orbits and cause them to slowly spiral into the Sun. If we assume that the meteoroid complex is in equilibrium, we can derive a model by assuming that these forces balance meteoroid production rates. Gr\"{u}n et al.~used this approach in combination with data from Pioneer 8 and 9, HEOS, the Pegasus satellites, and lunar microcraters to derive their famous 1985 model.

\begin{figure}
\centering
\includegraphics[width=\linewidth]{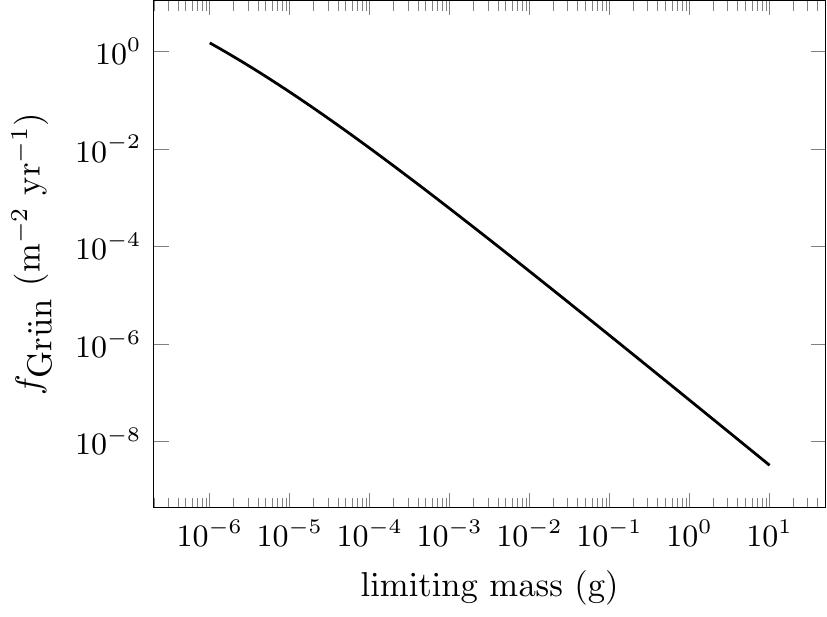}
\caption{Meteoroid flux at 1~au according to \protect\cite{1985Icar...62..244G}.}
\label{fig:fgrun}
\end{figure}

\begin{figure}
\centering
\includegraphics[width=\linewidth]{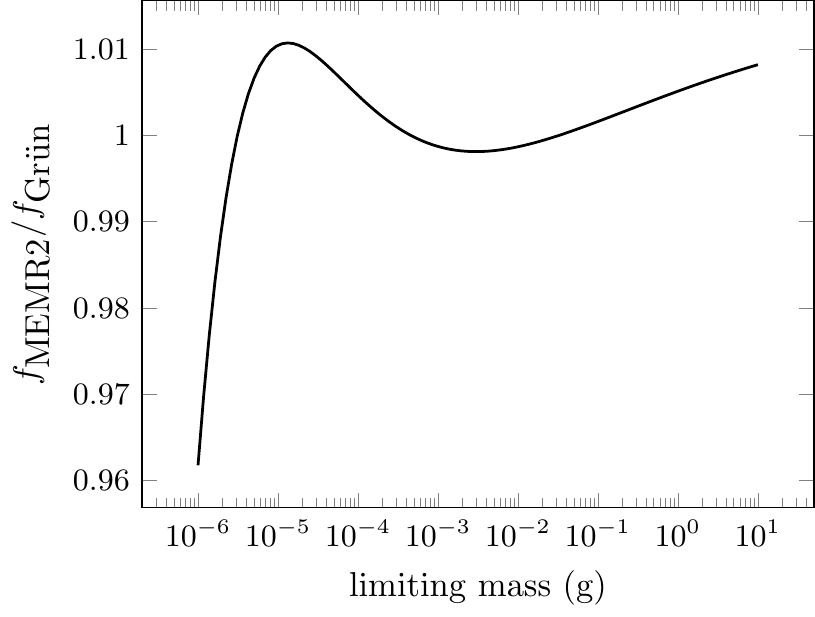}
\caption{Mass scaling function used by MEMR2 relative to that of Gr\"{u}n et al.~\protect\cite{1985Icar...62..244G}.}
\label{fig:mdif}
\end{figure}

We, like many other modelers, adopt the Gr\"{u}n et al.~model (or rather, the shape of this model) for our meteoroid mass distribution. The full analytic form of this distribution is given by Eq.~A3 of Gr\"{u}n et al.~\cite{1985Icar...62..244G} and is shown in Fig.~\ref{fig:fgrun}; the depicted flux is that on a randomly tumbling plate in interplanetary space at a heliocentric distance of 1~au. Previous versions of MEM used an approximation of the Gr\"{u}n et al.~model; MEM~3 replaces this approximation with the full analytic form. The difference is negligible (see Fig.~\ref{fig:mdif}). 

While MEM~3 uses the Gr\"{u}n et al.\ relation to scale the meteoroid flux to an arbitrary limiting mass, the amplitude of the flux is instead tied to the meteoroid flux observed at the top of the atmosphere by the Canadian Meteor Orbit Radar (CMOR) \cite{2008Icar..196..144C}. Equation~A3 of Gr\"{u}n et al.\ is applied to all flux components; the sources are assumed to follow the same mass distribution. As a result, the relative strength of the sporadic sources does not vary with limiting mass. However, studies have measured distinct mass indices for the sporadic sources \cite{2011MNRAS.412.2033B}, and we are considering adopting source-specific mass distributions in future versions of MEM. This would require replicating or replacing the orbital populations generated by \cite{Jones:2004uw}.

\subsection{Orbital populations}
\label{sec:orbits}

The sporadic meteoroid complex as observed at the Earth is organized into several ``sources.'' Each source describes a group of meteors that, like members of a meteor shower, have similar radiants (or directionality). Sporadic source radiants have a wider dispersion than shower radiants, however. They have a very wide range of velocities and are also present throughout the year, in contrast to showers, which tend to persist for only hours or weeks. When viewed in a Sun-centered ecliptic frame, the radiants of helion meteors are clustered around a point near the sunward direction, while antihelion meteors cluster around a point near the antisunward direction. The north and south apex sources are concentrated around points just north and south of the Earth's direction of motion around the Sun; the north and south toroidal sources lie further north and south of the ram direction. Table \ref{tab:sources} gives the approximate observed position of each sporadic source \cite{2008Icar..196..144C} and the angular center of the bin containing the greatest flux for each source as modeled using MEM~3.

\begin{table}
    \centering
    \begin{tabular}{ccccc}
        & \multicolumn{2}{c}{observed} 
        & \multicolumn{2}{c}{MEM~3} \\
        source & $\lambda_g - \lambda_\odot$ & $\beta_g$
               & $\lambda_g - \lambda_\odot$ & $\beta_g$ \\
        \hline
        helion & 340$^\circ$ & 0$^\circ$ & 
                 338.5$^\circ$ & 0$^\circ$ \\
        antihelion & 200$^\circ$ & 0$^\circ$ &
                201.5$^\circ$ & 0$^\circ$ \\
        north apex & 270$^\circ$ & +15--20$^\circ$ &
                270$^\circ$ & 12.5$^\circ$ \\
        south apex & 270$^\circ$ & -15--20$^\circ$ &
                270$^\circ$ & -12.5$^\circ$ \\
        north toroidal & 270$^\circ$ & +56$^\circ$ &
                270$^\circ$ & 54.5$^\circ$ \\
        south toroidal & 270$^\circ$ & -56$^\circ$ &
                270$^\circ$ & -54.5$^\circ$ 
    \end{tabular}
    \caption{Approximate location of the sporadic sources as observed and as modeled using MEM~3.}
    \label{tab:sources}
\end{table}

Earth-based meteor observations provide only a snapshot of the meteoroid environment at 1 au; extrapolating an interplanetary meteoroid environment from these observations requires modeling. MEM is based on the model of Jones \cite{Jones:2004uw}, who derived meteoroid orbits from parent body characteristics, parametrized various unknown quantities, and tuned those parameters so that the results matched sporadic meteor orbital distributions seen at Earth. To be more specific, these parent body populations were long-period comets (specifically those with orbital periods longer than 200 years), Halley-type comets (comets with periods between 20 and 200 years), Jupiter-family comets (comets with periods shorter than 20 years), and asteroids. Halley-type comets are sometimes grouped with other long-period comets, but we will follow Jones \cite{Jones:2004uw} in distinguishing between them. 

\begin{figure}
\centering
\includegraphics[width=\linewidth]{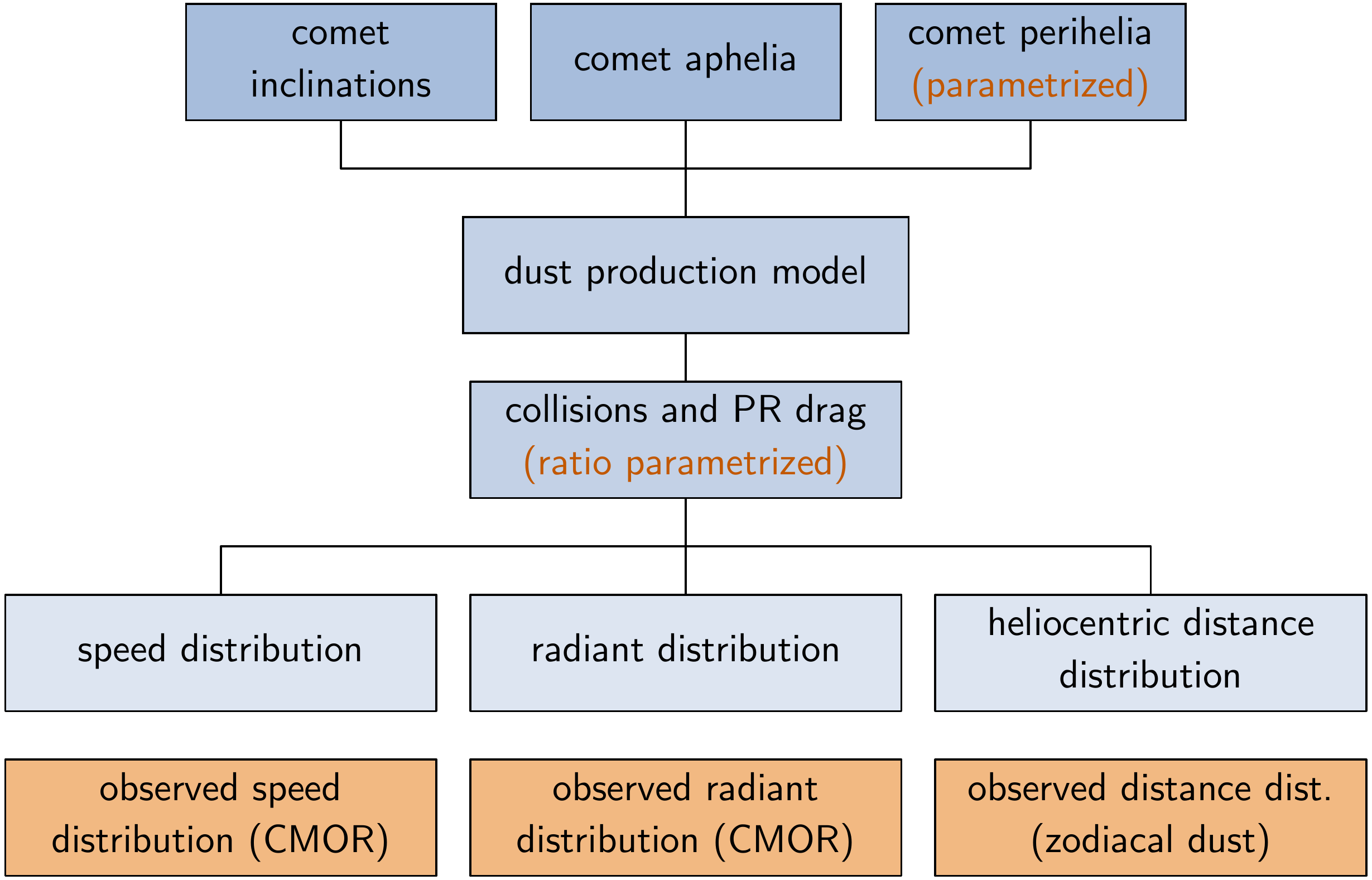}
\caption{Summary of the process used by \protect\cite{Jones:2004uw} to model the meteoroid environment. }
\label{fig:spormod}
\end{figure}

Jones developed a parametrized description of each parent body population's inclination, aphelion, and perihelion distributions. While the inclination and aphelion distributions were well-characterized, comet perihelion distributions are less well-known and Jones therefore developed an analytic form for each perihelion distribution whose parameters could be allowed to vary \cite{Jones:2004uw}. These distributions were then convolved with a dust production model and their evolution modeled, taking processes such as collisions and radiative forces into account. Once again, unknown quantities -- such as the ratio of the Poynting-Robertson (PR) timescale to the collisional timescale -- were parametrized. The evolved meteoroid distributions were then compared with the observed distribution of meteoroid speeds and radiants and the distance distribution from zodiacal dust measurements. These comparisons between the modeled and observed populations allowed Jones \cite{Jones:2004uw} to constrain his model parameters. His process is shown as a flow chart in Figure \ref{fig:spormod}. 

\begin{figure}
    \centering
    \includegraphics[width=\linewidth]{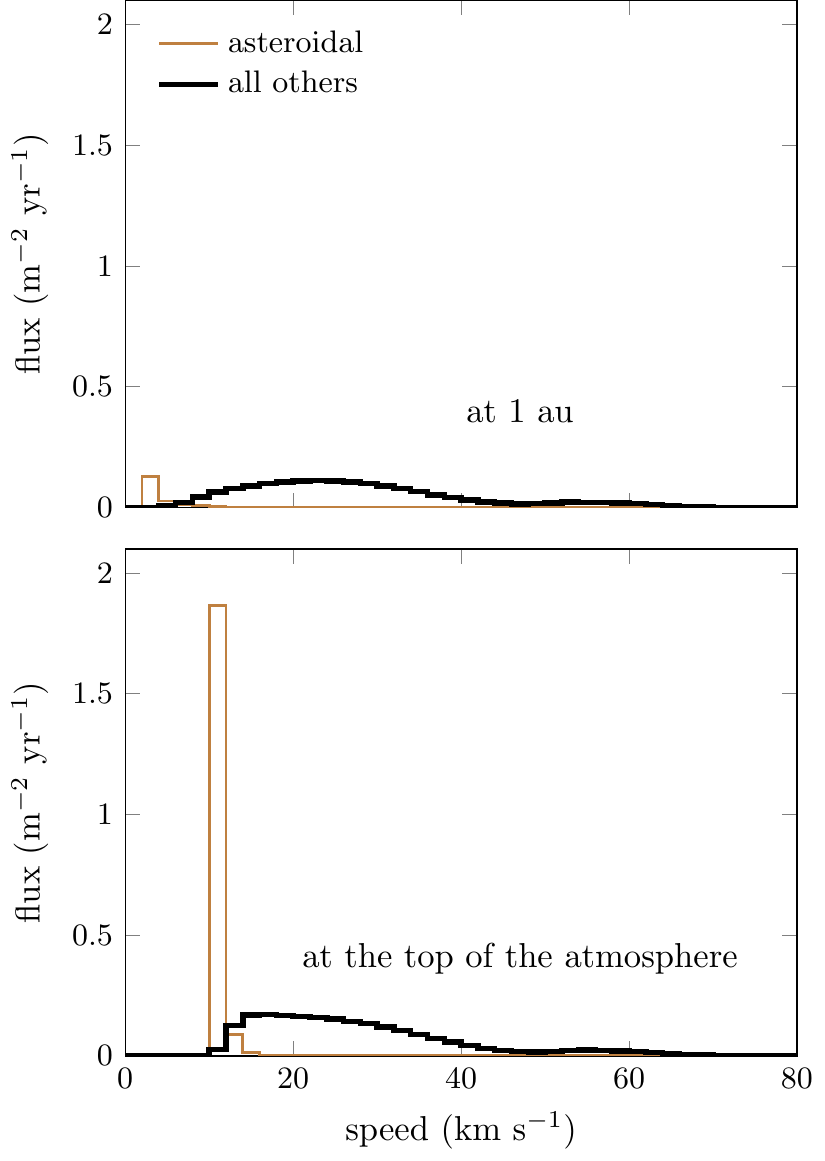}
    \caption{Flux and velocity of the asteroidal population compared to that of all other meteoroids.}
    \label{fig:ast}
\end{figure}

According to Jones \cite{Jones:2004uw}, the long-period comets can produce meteoroids similar to those in the apex sporadic sources, Halley-type comets can produce the toroidal sources, and Jupiter-family comets can produce the helion and antihelion sources. These linkages have been supported by a number of independent modeling efforts \cite{2009Icar..201..295W,2011ApJ...743..129N,2014ApJ...789...25P}. However, when Jones \cite{Jones:2004uw} modeled the evolution of particles ejected from asteroids, he obtained a set of radiants that matched no feature seen in the meteoroid environment seen at Earth. He concluded that meteoroids originating from asteroids must be a negligible component of the meteoroid environment. This is somewhat muddied by the fact that the asteroidal population produces very slow meteors, which are exceptionally difficult to detect \cite{2017P&SS..143...71S}, particularly using radar \cite{1997MNRAS.288..995J,2017P&SS..143..209M}. However, slow meteors also experience a large degree of gravitational focusing. We find that if we weight the asteroidal source so that it constitutes a modest 10\% of the flux in interplanetary space, this results in a flux at Earth that massively exceeds that seen at the top of the atmosphere. Figure \ref{fig:ast} shows the flux and velocity distribution of the asteroidal population (in brown) and that of all other modeled meteoroids (in black) at 1~au and at the top of the atmosphere. Gravitational focusing causes a large enhancement of the asteroidal component at the top of the atmosphere that is not seen in the data; we therefore eliminate it from MEM~3 entirely.

MEM~3 uses the remaining three populations to model the meteoroid environment. We have adjusted the relative strengths of these populations to better match observations from CMOR while maintaining the value of the total flux. The strength of the helion/antihelion source has been multiplied by a factor of 1.52, the apex source by a factor of 6.37, and the toroidal source by a factor of 0.339. As a result, the helion/antihelion source constitutes about 65\% of the flux at the top of the atmosphere, the apex source about 10\%, and the toroidal source about 25\%. 

MEM shares a number of features with other dynamical models. For instance, ref.\ \cite{2019ApJ...873L..16P} also reproduce the heliocentric dust density profile obtained from zodiacal light measurements \cite{1983A&A...118..345L}. The dominance of the Jupiter-family comet, or helion/antihelion, population in MEM is a feature shared by a number of models and studies besides our own \cite{2009Icar..201..295W,2011ApJ...743..129N,2018GeoRL..45.1713J,2019ApJ...873L..16P}. Some of these models do include small quantities of asteroidal meteoroids; for instance, \cite{2019JGRE..124..752P} includes an asteroidal component but finds their contribution at the Moon to be negligible and thus unconstrainable by LDEX. Acceptance of a dominant cometary population is not universal -- \cite{2012ApJ...749L..40C} argues that a higher contribution from asteroids better matches the abundance of 3He in ocean sediments -- but it is favored by most current dynamical models.

\begin{figure}
    \centering
    \includegraphics[width=\linewidth]{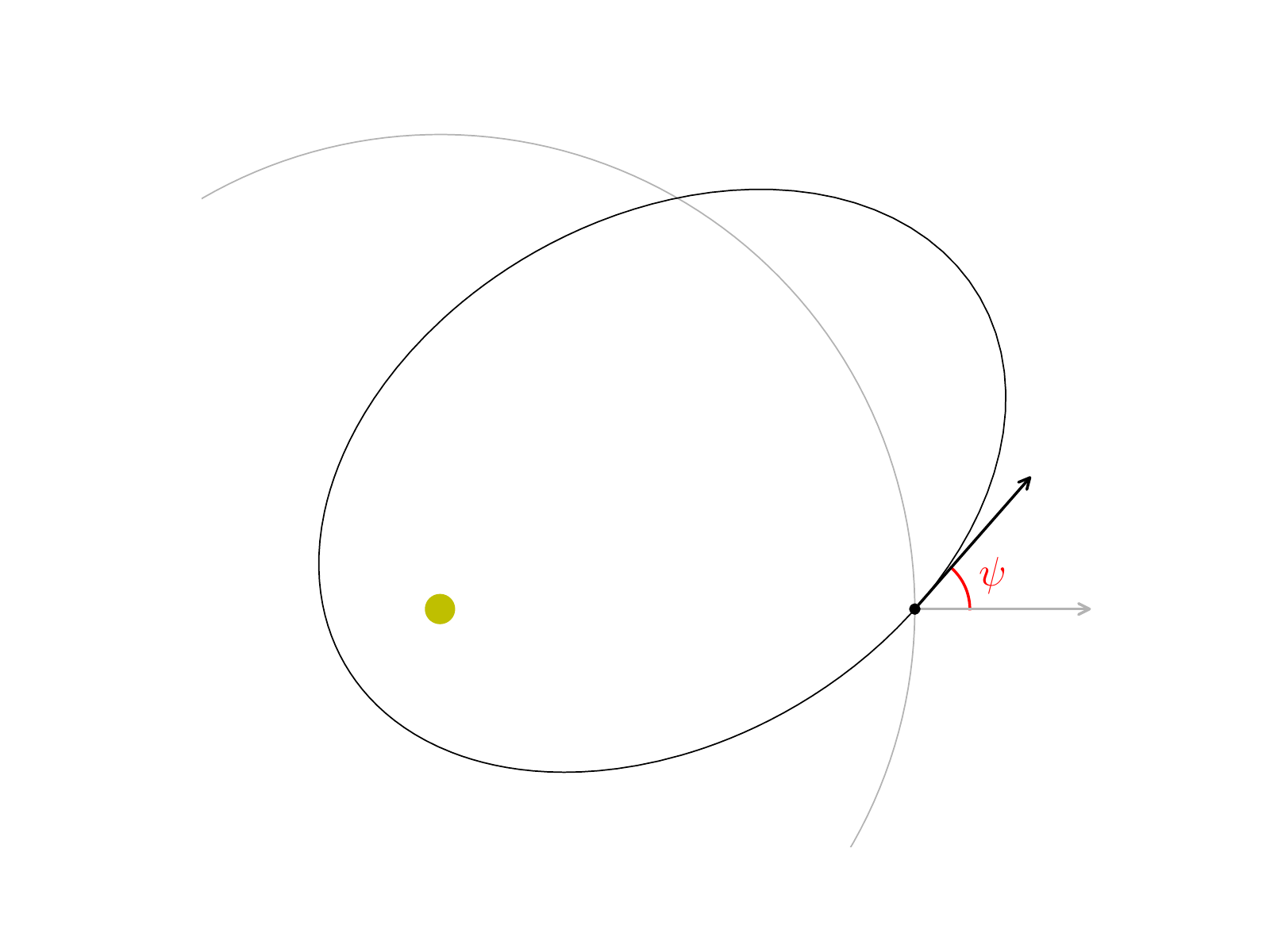}
    \caption{Diagram of the angle, $\psi$, that describes the non-radial component of the velocity vector.}
    \label{fig:psi}
\end{figure}

In practical terms, the primary output of Ref.~\cite{Jones:2004uw} is several sets of artificial meteoroid orbits (or rather, partial orbits) that are derived from meteoroid parent bodies and are tuned to match certain meteor observations. Each partial orbit is defined by three parameters: a semi-major axis, $a$; an eccentricity, $e$; and an inclination, $i$. Each population is assumed to have evenly distributed values of argument of pericenter, longitude of ascending node, and mean anomaly. The heliocentric ecliptic velocity of the meteoroid at heliocentric distance $r$ is 
\begin{align}
\vec{v} &= v \, 
	\begin{pmatrix} \pm \cos{\psi} \\
		\textcolor{white}{\pm} \sin{\psi} \cos{i} \\
		\pm \sin{\psi} \sin{i}
	 \end{pmatrix} \label{eq:vvec} \, , & \mbox{where} \\
v &= \sqrt{G M_\odot \left( 2/r - 1/a \right) } \, , & \mbox{and} \\
\psi &= \arcsin \left( \frac{a^2(1-e^2)}{r(2 a - r)} \right) \, ,
\end{align}
where $\psi$ describes the angle between the velocity vector and its radial component (see Figure \ref{fig:psi}).

There are thus four possible velocities corresponding to the meteoroid's values of $a$, $e$, and $i$ and the spacecraft's heliocentric distance $r$. MEMR2 randomly selected one out of the four possible trajectories for its calculation; MEM~3 allows the user to select either a low-fidelity mode, in which one of the four trajectories is randomly chosen, or a high-fidelity mode, in which all four trajectories are used. MEM~3 requires users to use the high-fidelity mode when calculating the standard deviation of the flux along their spacecraft's trajectory.

\subsection{Radial distribution}

As mentioned in the previous section, MEM assumes that the meteoroids within its populations have evenly distributed values of mean anomaly. As a result, the radial distribution of particles with a given $a$, $e$, and $i$ is inversely proportional to the radial velocity, and the probability of encountering the meteoroid at a particular heliocentric distance is therefore:
\begin{align}
    P(r) &= 
    \begin{dcases}
    \frac{r}{\pi a} 
        \frac{1}{\sqrt{(r-q)(Q-r)}} &
        \text{if } q \le r \le Q \\
    0 & \text{otherwise}
    \end{dcases}
\label{eq:rprob}
\end{align}
where $q = a(1-e)$ is perihelion and $Q = a(1+e)$ is aphelion. The above equation is only valid for values of $r$ between $q$ and $Q$.

\begin{figure}
    \centering
    \includegraphics[width=\linewidth]{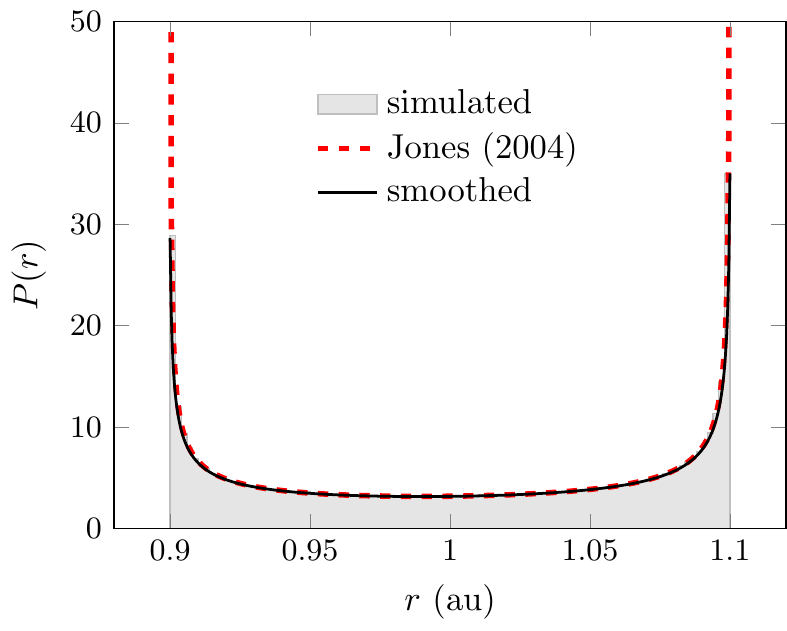}
    \caption{Probability distribution of heliocentric distance for an orbit with $a=1$~au and $e=0.1$.}
    \label{fig:rprob}
\end{figure}

\begin{figure}
    \centering
    \includegraphics[width=\linewidth]{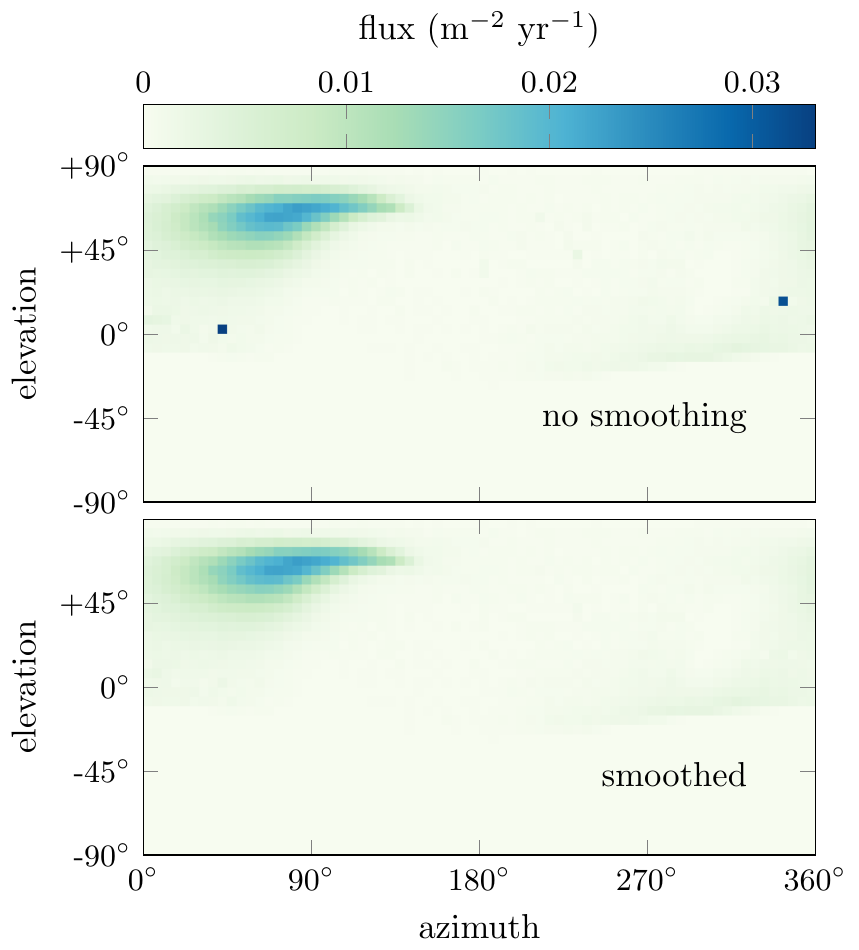}
    \caption{Example flux map with and without a smoothing factor included in the radial probability.}
    \label{fig:hotpix}
\end{figure}

Equation \ref{eq:rprob} has singularities at $r = q$ and $r = Q$ (see Figure \ref{fig:rprob}). As a result, whenever a spacecraft happens to travel close to one of the meteoroid pericenter values in MEM's orbital distributions, the result is a ``hot pixel'' in the flux map, where one meteoroid dominates the flux. The top panel of Figure \ref{fig:hotpix} shows an example of this effect; this flux map corresponds to a single International Space Station (ISS) state vector that happens to fall close to a particular meteoroid's perihelion. However, individual meteoroids are not tracked in the way that asteroids and comets are; the individual meteoroids in MEM are simply representatives of the modeled populations. Thus, there is no value in retaining the singularities and, in MEM~3, we have opted to smooth the radial probability over a fixed length scale of 0.01~au:
\begin{align}
    P(r) &= 
    \begin{dcases}
    \frac{r}{\pi a} 
            \frac{1}{\sqrt{(r-q)(Q-r) + 
            (\mbox{0.01 au})^2}} &
        \text{if } q \le r \le Q \\
    0 & \text{otherwise}
    \end{dcases}
\label{eq:rprob2}
\end{align}
This substantially reduces the effect of these singularities (see the bottom panel of Figure \ref{fig:hotpix}), but slightly reduces the total flux. We find that, on average, the flux is reduced by about 1\% due to this smoothing. Greater values of the smoothing length scale reduce the flux further; we found that our choice of 0.01~au was adequate to mitigate the singularities and we deemed the corresponding 1\% reduction in flux acceptable.

The reader may notice that Equations \ref{eq:rprob} and \ref{eq:rprob2} contain no dependence on inclination or distance from the ecliptic plane. Instead, the particle density is assumed to be independent of ecliptic latitude. To first order, this may be a reasonable approximation for Halley-type and long-period comets, which possess a wide range of eccentricities, but it is not a reasonable assumption for Jupiter-family comets \cite{Rickman2017}. Instead, one might expect the strength of the helion and antihelion source to diminish with increasing ecliptic latitude. For the rare spacecraft, such as \emph{Ulysses}, that ventures far from the ecliptic plane, MEM~3 likely overestimates the incident meteoroid flux. Additionally, the distribution of meteoroid inclinations intercepting the spacecraft is assumed to be independent of spacecraft location. Thus, the directionality at the boundaries of MEM's heliocentric distance range will be less accurate than it is for spacecraft near 1~au.

\subsection{Gravitational focusing and shielding}
\label{sec:planet}

Planets and other massive bodies bend and block meteoroid trajectories. The gravitational pull of a massive body accelerates and concentrates meteoroid trajectories in a phenomenon known as gravitational focusing. As a result, spacecraft in low-altitude orbits will experience a higher meteoroid flux, on average, and impacts will tend to occur at higher velocities than they would in interplanetary space. This is countered to some extent by planetary shielding, in which the planet or other large body physically blocks some meteoroids from reaching the spacecraft. Depending on the mass and size of the central body and its proximity to the spacecraft, either gravitational focusing or shielding may dominate.

\begin{figure*}
\centering
\includegraphics{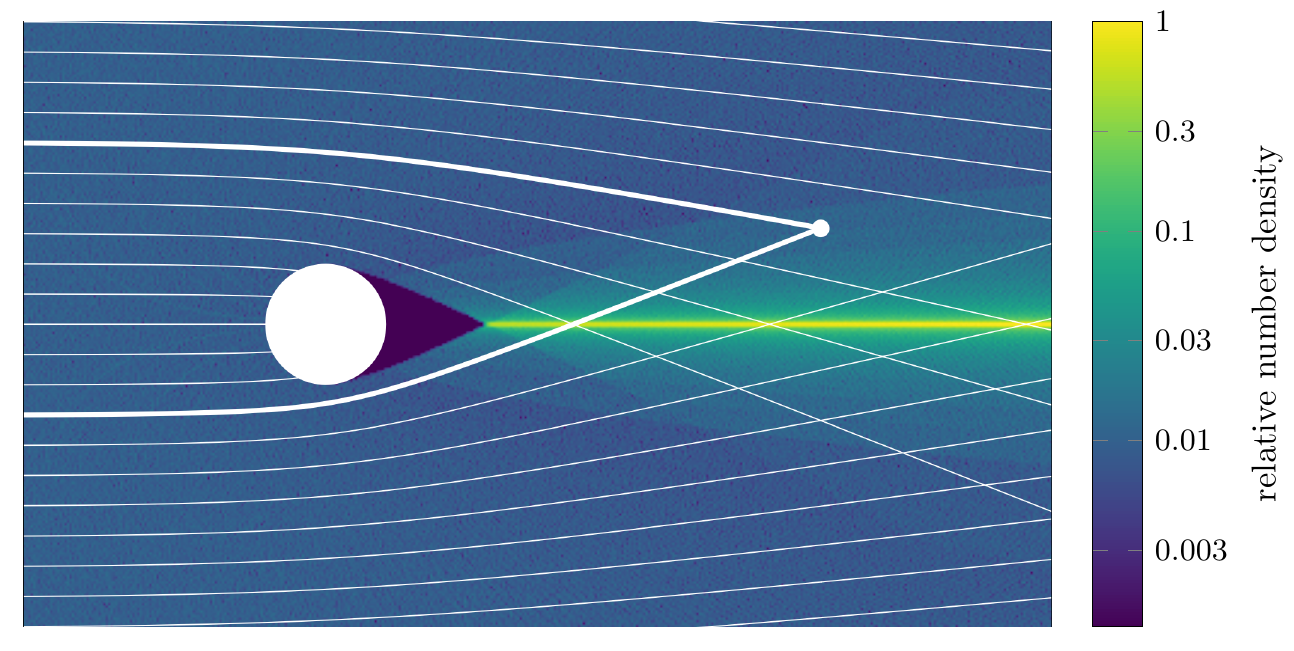}
\caption{Illustration of how meteoroid trajectories (lines) and number density (heat map) are altered near a massive body.}
\label{fig:grav}
\end{figure*}

The effects of gravitational focusing and shielding on the meteoroid environment vary with angle, as illustrated in Figure \ref{fig:grav}.  This example corresponds to a perfectly collimated meteor stream with an interplanetary speed of 15~km~s$^{-1}$; all meteoroids pass from left to right in this plot. The color scale represents the spatial density of meteoroids, while the white lines follow individual meteoroid trajectories. The white circle represents the Earth (including 100~km of atmosphere). Note that Figure~\ref{fig:grav} depicts the effects of gravitational focusing on a single meteoroid radiant (i.e., initial direction of motion). These effects must be applied to each meteoroid radiant to obtain the full focused and shielded sporadic environment.

As illustrated in Fig.~\ref{fig:grav}, two spacecraft at the same altitude above the same massive body may experience different flux levels if their angular position relative to the directionality of the meteoroid environment differs. A full set of analytic expressions governing these effects are provided by \cite{1997AdSpR..19..301S}, \cite{2002dsso.conf..359M}, and \cite{2007MNRAS.375..925J}. MEM follows \cite{1997AdSpR..19..301S} in applying gravitational focusing and shielding; \cite{1997AdSpR..19..301S} is based on earlier, unpublished work by Neil Divine \cite{DivineMemo}. MEM does not include second-order gravitational focusing in which meteoroid paths are consecutively focused by two different bodies, such as the Earth and the Moon.

As is apparent in Figure \ref{fig:grav}, the overall effect of gravitational focusing and shielding varies with location. However, the effect of gravitational focusing on the total flux incident on a sphere surrounding the Earth follows a fairly simple relation:
\begin{align}
\bar{\eta}_{g} &= 1 + \frac{v_{esc}^2}{v_i^2} \, , \label{eq:opik}
\end{align}
where $\bar{\eta}_{g}$ is the factor by which the flux is increased due to gravitational focusing, averaged over the surface of the sphere; $v_{esc}$ is the escape velocity at the surface of the sphere; and $v_i$ is the initial speed of the meteoroid relative to the Earth, before gravitational focusing is applied. This equation may be used independently to verify that the gravitational focusing and shielding algorithms are correct; \cite{2007MNRAS.375..925J} terms this the ``\"{O}pik test."

\begin{figure} 
\centering
\includegraphics[width=0.8\columnwidth]{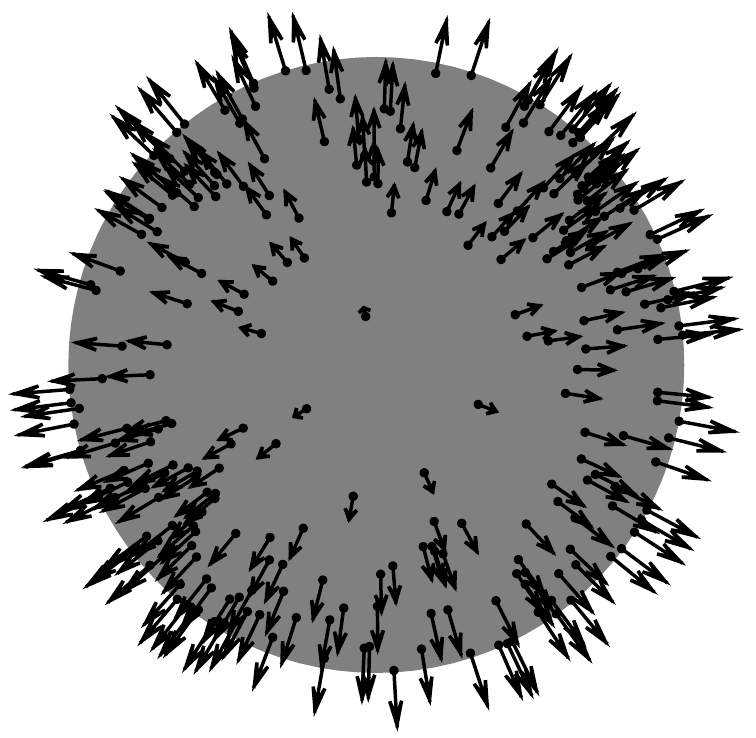}
\caption{Diagram of state vectors used for the ``\"{O}pik test'' of gravitational focusing.}
\label{fig:svtest}
\end{figure}

We apply the \"{O}pik test to MEM~3 as follows. We construct two sets of state vectors, one corresponding to a sphere of radius $R_\oplus + 100$~km, and the other to a sphere of radius $R_\oplus + 100$,000~km. The positions of the state vectors within each set are isotropically distributed on the surface of the sphere (see Figure \ref{fig:svtest}). Each state vector is given a velocity that points directly outward but has a magnitude of only $10^{-6}$~km~s$^{-1}$. These state vectors do not correspond to any physical object; instead, they are designed to take advantage of MEM's ``body-fixed'' coordinate frame to probe the flux incident on the sphere. The ``ram'' direction will always point outward, and thus the average ``ram'' flux is equivalent to the average flux on the surface of the sphere.

At an altitude of 100,000~km, gravitational focusing will be fairly minimal; at 100~km, it should be significant. We can therefore use the total flux on the sphere at 100,000~km, convolve the speed distribution at that altitude with Equation \ref{eq:opik}, and predict the equivalent flux values at an altitude of 100~km. We will also need to convert each speed at 100,000~km to its equivalent speed at 100~km:
\begin{align}
v_1 = \sqrt{v_2^2 - \frac{2 G M_\oplus}{R_\oplus + h_2} 
+ \frac{2 G M_\oplus}{R_\oplus + h_1}} \, , \label{eq:vshift}
\end{align}
where $v_1$ is the speed at an altitude of $h_1 = 100$~km, and $v_2$ is the speed at $h_2 = 100$,000~km. The speed bins of our distribution will be unevenly compressed, and we will therefore need to divide by the bin width to maintain units of m$^{-2}$~yr$^{-1}$~(km~s$^{-1}$)$^{-1}$ in the focused velocity distribution. Finally, the distribution of speeds \emph{within} the bins cannot be determined from MEM's output files, and will introduce a certain amount of uncertainty in the results.

When conducting this test on MEMR2, we found that the results appeared to be ``overfocused.'' MEMR2 modified the Divine gravitational focusing model by multiplying Equation 7 of \cite{1997AdSpR..19..301S} by a factor of $v_f/v_i$, where $v_i$ is the speed of the meteoroid before focusing and $v_f$ is the speed after focusing. This factor was applied in order to bring Equation 7 of \cite{1997AdSpR..19..301S} into agreement with other works; e.g., \cite{2002dsso.conf..359M}. However, Equation 7 of \cite{1997AdSpR..19..301S} provides a \emph{number density} gravitational enhancement factor, not a flux enhancement factor. The conversion of Equation 7 to a flux enhancement factor, when combined with Equation 14 of \cite{1997AdSpR..19..301S}, results in erroneously high gravitational focusing. This misunderstanding is likely due to the unclear wording in \cite{1997AdSpR..19..301S}, which mentions flux but not number density; the same misinterpretations seems to appear in \cite{2007MNRAS.375..925J}. The factor of $v_f/v_i$ has been removed from the number density enhancement factor in MEM~3.

Our review also revealed that MEMR2 was only using one of two possible trajectories by which gravitationally focused meteoroids can reach a spacecraft. As discussed by \cite{Jones:2004uw} and shown in Figure \ref{fig:grav}, there are often two paths that contribute to the flux at the spacecraft's location, a so-called ``long'' and ``short'' path. The long path makes a sharper bend around the far side of the planet, which requires it to pass closer to the planet's center of mass. The long path is often blocked by planetary shielding, but not always. It makes a small contribution to the flux seen by the spacecraft and should be included; we include both gravitationally focused paths in MEM~3.

\begin{figure}
\centering
\includegraphics[width=\linewidth]{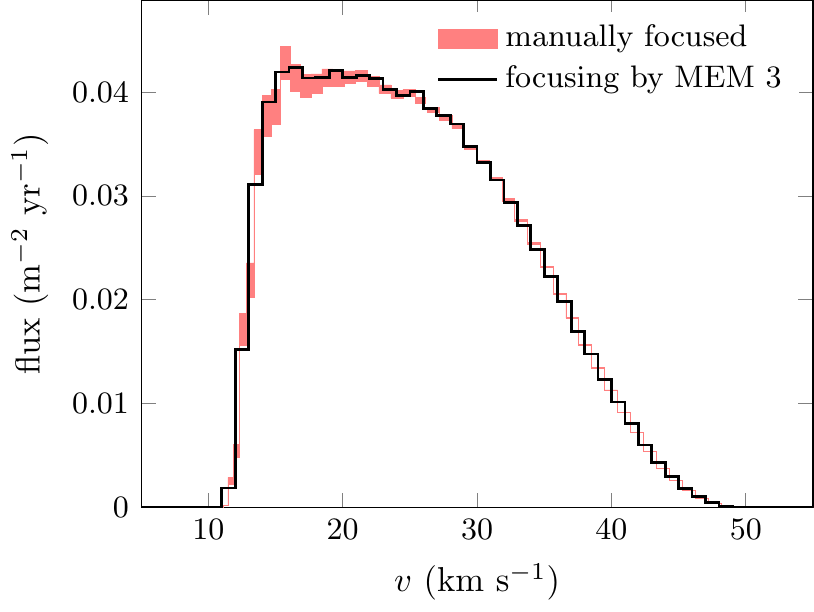}
\caption{
Helion meteoroid flux on the top of the atmosphere as computed by MEM~3.
}
\label{fig:opiktest}
\end{figure}

Figure~\ref{fig:opiktest} compares two (helion) speed distributions. The distribution shown in black corresponds to a set of state vectors scattered randomly across the top of the atmosphere; MEM~3 computes the effects of Earth's gravitational focusing on flux and speed. The second distribution, shown in red, corresponds to a set of state vectors scattered randomly across a sphere of radius 100,000~km; these state vectors were used to obtain the near-Earth meteoroid environment under the influence of very little gravitational focusing, which was then adjusted using Equations \ref{eq:opik} and \ref{eq:vshift}. The uncertainty arising from the finite velocity bin width is reflected in the width of the red distribution. The two agree to within these uncertainties.

The Earth's atmosphere is also now included in calculating planetary shielding (this was omitted in MEMR2). Meteoroids begin to ablate about 100~km above the surface of the Earth, and thus the shielding radius of the Earth is taken to be 6471~km.
Mars is assumed to have 90~km of meteoroid-blocking atmosphere \cite{1996Icar..119..144A} and Venus is assumed to have 120~km of atmosphere \cite{2004Icar..168...23C,2006Icar..180....8M}.

\subsection{Density distribution}
\label{sec:density}

MEM, like many meteoroid models, historically assumed a single meteoroid bulk density of 1000 kg~m$^{-3}$. Other authors have sometimes assumed a single value of 2500~kg~m$^{-3}$ \cite{1985Icar...62..244G,1993Sci...262..550L}. Meteoroids are unlikely to possess a uniform density, however, and thus one of the primary goals for MEM~3 was to introduce a more realistic bulk density model.

\begin{figure*} \centering
\includegraphics[width=\linewidth]{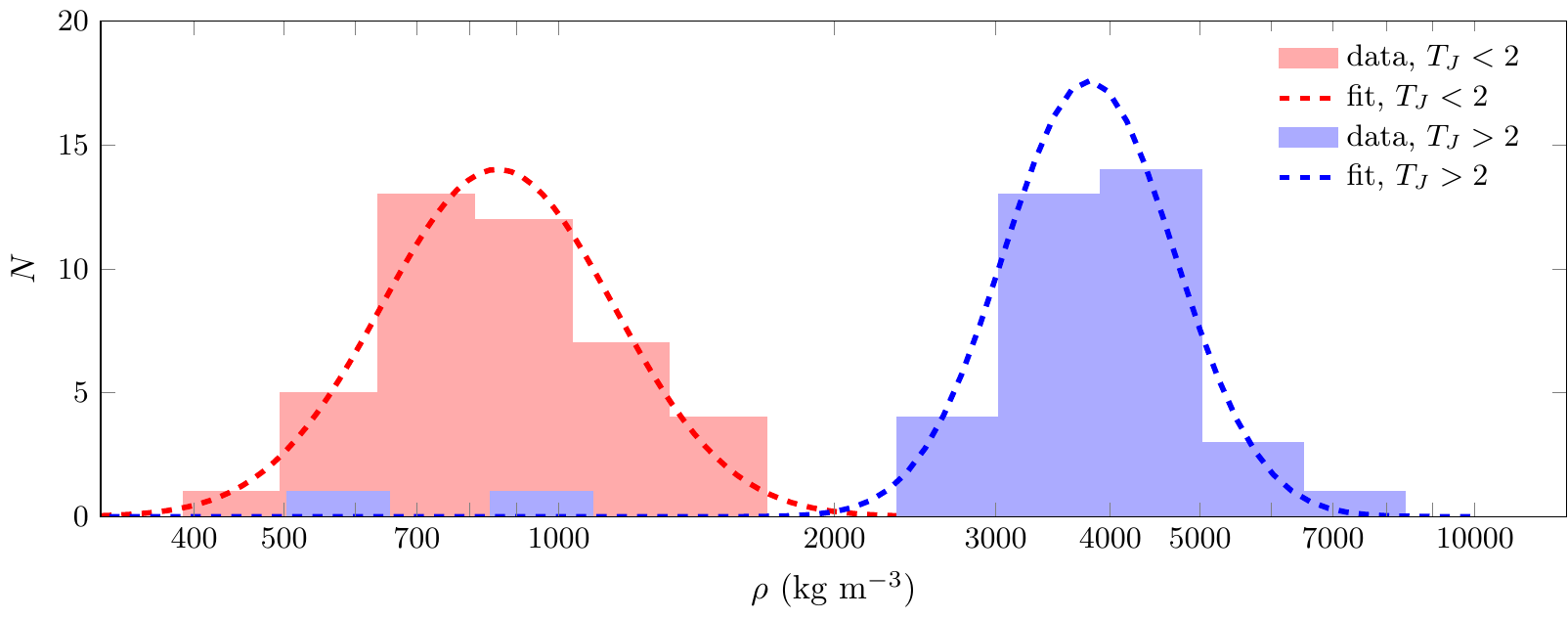}
\caption{Distribution of bulk densities for sporadic meteoroids }
\label{fig:density}
\end{figure*}

Kikwaya et al.~\cite{2011A&A...530A.113K} measured the densities of nearly 100 meteoroids by fitting the meteor ablation model of \cite{2004A&A...418..751C} to meteor light curves and trajectories. The results showed a correlation with Tisserand parameter, a combination of orbital elements that is often used to classify Solar System orbits as ``asteroidal'' or ``cometary". Further analysis by \cite{2017MNRAS.472.3833M} demonstrated that the traditional measure of meteoroid physical properties, or ``Ceplecha type'' \cite{1966BAICz..17..347C}, displayed no correlation with the density values obtained by \cite{2011A&A...530A.113K}. As a result, \cite{2017MNRAS.472.3833M} used the density measurements of \cite{2011A&A...530A.113K} and an earlier work \cite{2009A&A...497..851K} to derive a bimodal sporadic meteoroid bulk density distribution that could be incorporated into models such as MEM (see Fig.~\ref{fig:density}).
Sporadic meteoroids with orbits like those of long-period comets -- i.e., those with Tisserand parameters, $T_J$, less than 2 -- have lower densities (see red area of Fig.~\ref{fig:density}). In contrast, those with orbits more similar to asteroids or short-period comets -- i.e., those with $T_J > 2$ -- have higher densities (see blue area of Fig.~\ref{fig:density}). The density values represented by the histograms are taken from \protect\cite{2011A&A...530A.113K} and \protect\cite{2009A&A...497..851K}.  Reference \cite{2017MNRAS.472.3833M} fit a normal distribution to each group (dashed lines), where the probability $P$ of drawing (the log-base-10 of) a particular density value is:
\begin{align}
    P(\log_{10} \rho) &= \frac{1}{\sqrt{2 \pi \sigma^2}}
        e^{-(\log_{10} \rho - \mu)^2/2\sigma^2}
    \label{eq:logrho}
\end{align}
where $\rho$ is density. For the low-density population, $\mu = 2.933$ and $\sigma = 0.127$; for the high-density population, $\mu = 3.579$ and $\sigma = 0.093$. These values correspond to densities of 857~kg~m$^{-3}$ and 3792~kg~m$^{-3}$ with standard deviations of 34\% and 24\%, respectively. Reference \cite{2017MNRAS.472.3833M} contains additional information regarding the data selection, fitting process, and comparison with other works.

To accommodate this bimodal density distribution, MEM now generates two sets of output files that correspond to the low and high density populations. The low density population incorporates both the apex and toroidal sources, while the high density population consists of the helion/antihelion source. Each set of output files is accompanied by a density distribution file; per set of files, the density distribution can be taken as independent of radiant and direction. The number of output files has therefore slightly more than doubled since MEMR2, but we have avoided adding an additional density dimension to each file.

\subsection{Velocity distribution}

\begin{figure}
    \centering
    \includegraphics[width=\linewidth]{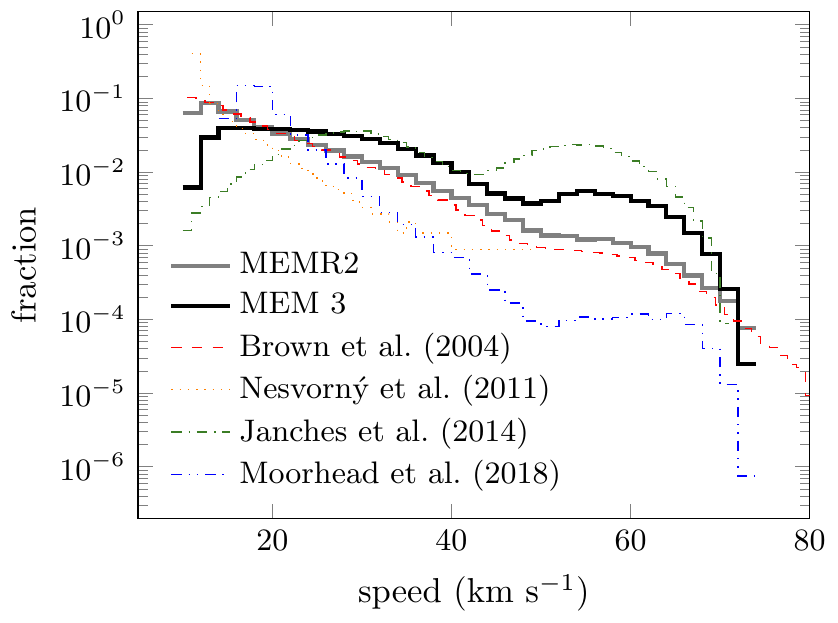}
    \caption{Top-of-atmosphere speed distribution from MEMR2, MEM~3, and other recent works.}
    \label{fig:vdist}
\end{figure}

MEMR2 introduced an adjusted speed distribution that was intended to bring the predicted top-of-atmosphere speed distribution in line with that of \cite{2004EM&P...95..617B}. This was done by calculating every meteoroid velocity at the top of the atmosphere as well at the spacecraft's location and applying a reweighting factor based on the top-of-atmosphere speed. This was implemented only in the Earth sub-module of MEMR2; the Earth sub-module thus did not have a speed distribution that was consistent with the interplanetary and lunar submodules. Meanwhile, a more recent analysis of the meteoroid speed distribution was conducted using the same data \cite{2017P&SS..143..209M} but debiased using a more modern treatment of the ionization efficiency \cite{1997MNRAS.288..995J,2016GeoRL..43.3645T}. This new speed distribution contains fewer fast meteors than that of \cite{2004EM&P...95..617B} (see Fig.~\ref{fig:vdist}).

Re-weighting the speed distribution to match observations essentially overrides the physics-based approach of Jones \cite{Jones:2004uw} and replaces it with an empirical one. As the current developers of MEM favor a physics-based approach, and the evidence in favor of the \cite{2004EM&P...95..617B} speed distribution has been weakened, we have opted to remove the speed re-weighting in MEM~3.  An attempt was made to re-weight the source populations strengths to better match the speed distribution of \cite{2017P&SS..143..209M}; this approach would preserve the physics built into each individual source population. However, we were unable to find a new relative weighting that matched both the observed speed distribution and the observed radiant distribution (see Fig.~\ref{fig:vdist}), and therefore have opted to tailor our source strengths to match the radiant distribution measured by \cite{2008Icar..196..144C}.

The change in the resulting speed distribution is significant; it is approximately an order of magnitude different in some speed bins. However, this change is still small compared to the range in meteoroid velocity distributions obtained from either meteor observations or other models (refs. \cite{2004EM&P...95..617B,2011ApJ...743..129N,2014ApJ...796...41J,2018M&PS...53.1292M}, represented as the thin patterned lines in Figure~\ref{fig:vdist}). This list is not exhaustive -- there are numerous other measurements and models of the speed distribution in the literature. Thus, the change in the speed distribution between MEMR2 and MEM~3 lies within the current level of uncertainty. Additional data and analysis are needed to better constrain the speed distribution.

MEM~3 does offer finer resolution velocity binning than previous versions. MEMR2 reported fluxes in bins of 2, 4, or 5~km~s$^{-1}$ in width, while MEM~3 offers 1 and 2~km~s$^{-1}$ velocity binning. The larger bin widths were deemed to be too coarse -- a factor of 5 in velocity corresponds to about a factor of about 4 in cratering rate -- and thus have been removed from the set of user options. 

\begin{figure*}
    \centering
    \includegraphics{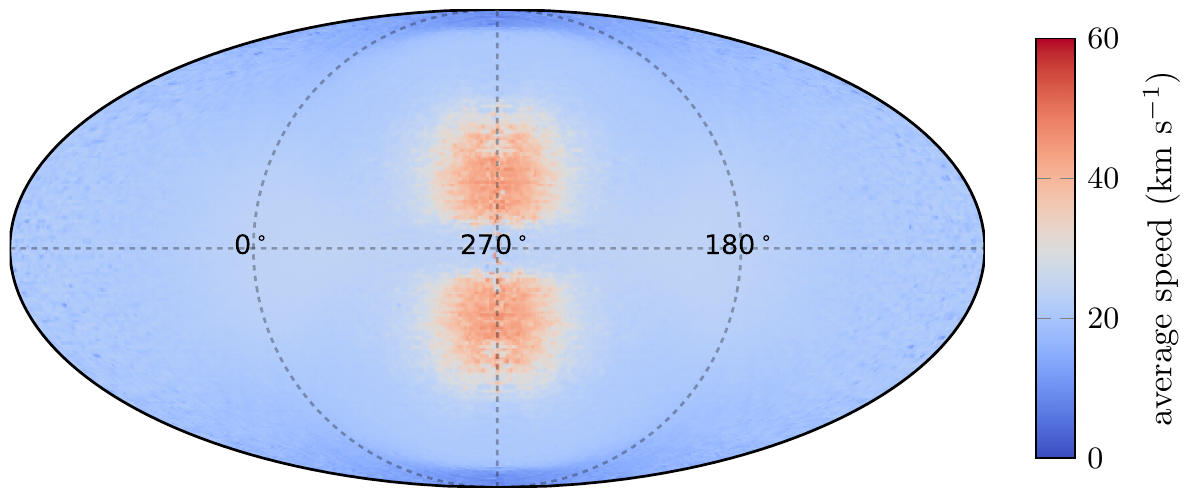} \\
    \includegraphics{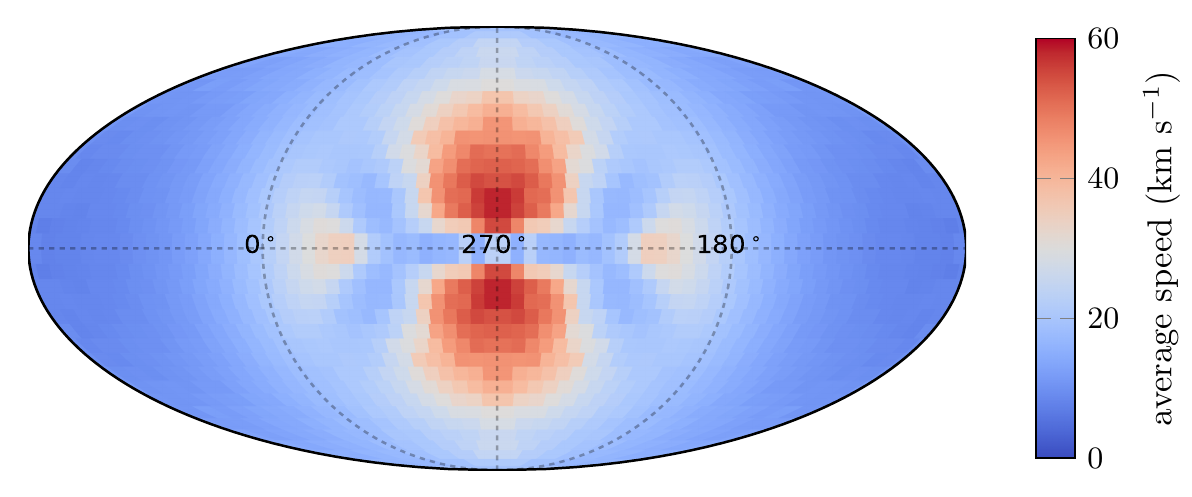}
    \caption{Average velocity per Sun-centered ecliptic radiant bin at 1~au from MEMR2 (top) and MEM~3 (bottom).}
    \label{fig:vrad}
\end{figure*}

The speed distribution is not independent of the directionality; rather, each radiant-speed pair must correspond to an orbit drawn from the Jones distributions \cite{Jones:2004uw}. In our review, we found that MEMR2 generated the correct speed distribution and radiant distribution for each source population, wrote these distributions to file, and then multiplied the two together to obtain the flux as a function of speed \emph{and} radiant. This is both inefficient and incorrect; in short:
\begin{align}
\frac{d}{d \theta} \frac{d}{d \phi} \frac{d F}{d v} &\ne \frac{1}{F} \frac{d F}{d \theta} \times \frac{1}{F} \frac{d F}{d \phi} \times \frac{d F}{d v} \, .
\end{align}
We have removed this assumption from MEM~3, preserving the physical correlations between meteoroid radiant and speed. This affects the speed-radiant maps and the average speed of meteoroids incident on each spacecraft surface. Figure \ref{fig:vrad} shows the speed-radiant map from both MEMR2 (top panel) and MEM~3 (bottom panel). MEMR2 ignores the correlations between radiant and density per source, substantially weakening the correlation between speed and direction, while MEM~3 fully preserves these correlations.

\subsection{Ephemerides}

MEM~3 breaks from its predecessors in discarding the submodule approach to handling spacecraft trajectories in different regions of space. MEMR2 required users to select an Earth, lunar, or interplanetary submodule to run depending on whether the spacecraft was in near-Earth, cis-lunar, or interplanetary space. These mutually exclusive submodules required users to divide transfer trajectories into pieces and run them separately. This approach was necessary due to the relatively low-precision ephemerides used by MEMR2 for the Earth and Moon \cite{1998aalg.book.....M}; these ephemerides were adequate to compute angles between the Earth, spacecraft, and meteoroid radiants within the Earth sub-module, for instance, but could not be used to convert a heliocentric spacecraft state vector to a geocentric state vector for spacecraft orbiting the Earth.

We have incorporated a more accurate ephemeris file into MEM~3; specifically, we use the JPL DE430 ephemeris.\footnote{https://naif.jpl.nasa.gov/pub/naif/generic\_kernels/spk/planets/aareadme\_de430-de431.txt (retrieved 23 May 2019)} This ephemeris is the current standard ephemeris for analyzing objects (particularly the moon) in the modern era. DE430 reports the positions of the planets with sub-kilometer accuracy; for more details, see \cite{2014IPNPR.196C...1F}. MEM introduces errors larger than 1~km by approximating the Earth and the Moon as spheres, and we therefore consider the DE430 completely adequate for transforming heliocentric state vectors to planetocentric state vectors and vice versa. Because these transformations are performed internally by MEM~3, users can input their state vectors relative to any major body in the inner Solar System, and the code will automatically detect and account for any nearby massive bodies. Users now also have the freedom to input state vectors and output results in either an ecliptic or Earth-equatorial reference frame, regardless of what body the spacecraft is orbiting. Our ephemeris routines are based on Piotr Dybczy\'{n}ski's translation of the JPL fortran code to C.\footnote{https://apollo.astro.amu.edu.pl/PAD/pmwiki.php?n=Dybol.JPLEph (retrieved 23 May 2019)}

\begin{figure*} 
    \centering
    \includegraphics[width=\linewidth]{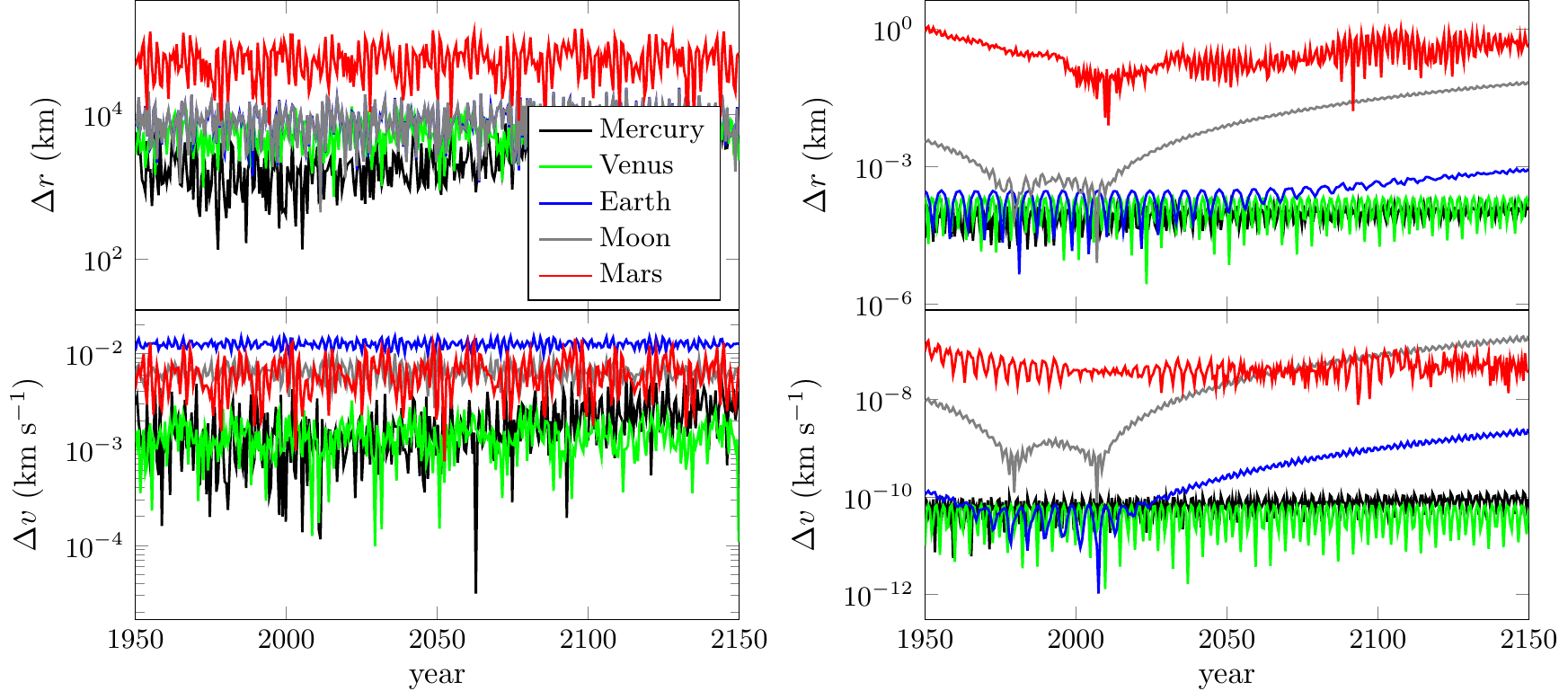}
    \caption{Difference in position and velocity calculated by JPL HORIZONS and MEMR2 (left) or MEM~3 (right).}
    \label{fig:dephem}
\end{figure*}

We tested both ephemeris routines against the JPL HORIZONS online ephemeris system,\footnote{https://ssd.jpl.nasa.gov/horizons.cgi (retrieved 23 May 2019)} which uses the DE431x ephemeris. The DE431 ephemeris has a longer time span and a less detailed lunar dynamical model; the DE430 and DE431 ephemerides should nevertheless produce very similar results, with differences in the positions of the planets less than 1 m between the two models, and differences in the position of the Moon of less than 10~m. The results of our comparison are shown in Figure~\ref{fig:dephem}.  We see that while the previous code was accurate to a few Earth radii (6371 km) in position and a few m s$^{-1}$ in speed, the new code is accurate to 1~km in position and 0.1~mm~s$^{-1}$ in speed. The planet Mars shows the largest differences between the two ephemerides, reaching a difference of about 1~km between the code and JPL Horizons. 
Thus, the new ephemeris routines offer at least five orders of magnitude improvement in precision, and the uncertainties are insignificant in the context of modeling the meteoroid environment.

As a result of this ephemeris update, MEM~3 is also capable of modeling the meteoroid environment seen by spacecraft in orbit around Mars, Mercury, and Venus in addition to those in orbit around the Earth, Moon, and Sun.  We do not, however, model the environment near minor planets, comets, or other special environments; such cases would require a separate, specialized approach. MEM~3 accepts spacecraft positions with heliocentric distances between 0.2~au and 2~au; extensive additional dynamical modeling is needed to extend MEM beyond this range.

The range of meteoroid speeds depends on heliocentric distance; meteoroids impact Mercury with relative speeds as high as 135~km~s$^{-1}$, while at Mars, meteoroids impact the atmosphere at speeds below 63~km~s$^{-1}$. The relative speeds of meteoroids impacting spacecraft orbiting these bodies may be even higher. To reduce the size of the output files as much as possible, MEM~3 preprocesses the spacecraft trajectory input file to determine the maximum possible meteoroid speed relative to the spacecraft, and uses this value to determine the size of the velocity array. The size of the output files will therefore vary in size depending on the input trajectory.

\subsection{Run time}
\label{sec:runtime}

MEM~3 offers a significant improvement in speed over its predecessor. For the same choice of fidelity, MEM~3 runs approximately three times as fast as MEMR2. Users may choose to sacrifice this improvement in run-time for an improvement in fidelity and receive results that have a four-fold improvement in fidelity in approximately the same run time as MEMR2.

This increase in speed has been obtained in part by avoiding the use of intermediate output files. Prior versions of MEM generated files describing the environment corresponding to each state vector in a spacecraft trajectory file. These files were then read back in by the code to calculate the average and standard deviation of the environment over the entire trajectory. This required a sometimes large amount of disk space and introduced rounding errors in the results. To avoid this behavior, MEM~3 calculates both the average and standard deviation of the flux in a single pass using the Welford algorithm
\cite{Welford:1962eb}. The flux-weighted mean and standard deviation of the average velocity on each surface of a cubic spacecraft is computed in a single pass using the West algorithm \cite{West1979}. Intermediate files are only written if the user specifically requests them, and are never read back in by the code.

\begin{figure} \centering
\includegraphics[width=\linewidth]{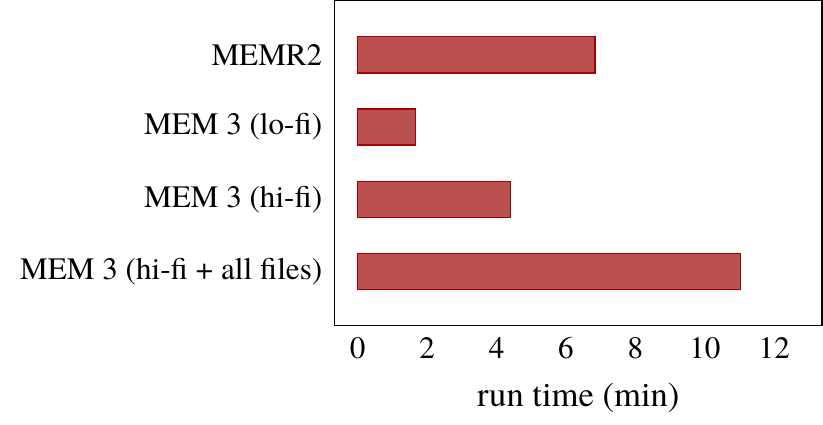}
\caption{Sample MEMR2 and MEM~3 execution times for various run types.}
\label{fig:runtime}
\end{figure}

The exact difference in run time between MEMR2 and MEM~3 depends on a number of factors, including the heliocentric distance of the spacecraft, resolution choices, number of output files produced, and the level of fidelity. When the user chooses to do a high-fidelity run and output all possible file types, MEM~3 run times can exceed those of MEMR2. Figure \ref{fig:runtime}) shows run times for a spacecraft in LEO that has been analyzed using MEMR2, MEM~3 in low-fidelity mode with minimal output, MEM~3 in high-fidelity mode with minimal output, and MEM~3 in high-fidelity mode with maximal output. All runs shown have an angular resolution of 1$^\circ$ and a velocity resolution of 2~km~s$^{-1}$.

\subsection{Other improvements}
\label{sec:improve}

In general, MEM~3 gives the user increased control over many aspects of the code, such as choosing which files to output, where to place output, what coordinate system to use, and so forth. For instance, the input file need not be located in the same directory as the executable; an absolute path may selected or specified. Similarly, the output files are placed in the user's directory of choice. Allowing users to select the location of their output files should avoid a common problem with MEMR2 installations, which is that the code by default attempted to save files to a directory to which typical users did not have write permissions, preventing the code from running. 

Users may select two different coordinate systems for their input and output, and the choice of axes (ecliptic or equatorial) is not determined by the choice of origin. The full suite of run choices are described in detail in the user manual.

In some cases, options have been removed; one example is the removal of the coarsest velocity resolution choices. We have also removed the lunar coordinate system, as it was never correctly implemented. MEMR2 treated the lunar coordinate system as a simple rotation of the ecliptic coordinate system by 1.5424$^\circ$; however, this does not bring the $x$-$y$ plane into alignment with the lunar equator. Users now can specify distances from the Moon in either an ecliptic or Earth-equatorial frame. We have also removed the ability to specify spacecraft trajectories using two-line element sets, or TLEs. MEMR2 converted each TLE into a single state vector, usually leading to a severe under-sampling of the orbit in question. Adequate sampling of an orbit will depend on the particular orbital elements; thus, rather than try to anticipate all possible types of orbits that could be input into MEM, we instead require users to use a tool such as STK to convert their TLEs to state vectors.

MEM~3 consists of a command-line executable that has been compiled for multiple operating systems, including at least one version of Windows, Mac OS X, and Linux. This executable may be run directly in a command-line environment or Windows users may opt to use a separate GUI program to assist them in running the code. The code is not parallelized and will not take advantage of multiple processors.

Finally, MEM~3 contains various internal improvements that will not be readily apparent to most users. All flux binning and averaging is now done internally, without the use of intermediate files, reducing rounding errors. Similarly, we have included additional digits in constants such as $\pi$. When using a random subset of state vectors, selection is now done \emph{without} replacement, maximizing the number of input state vectors used. We have also corrected a bug in the handling of 2-state-vector files. When the input files contained only two state vectors and the standard deviation was turned on, MEMR2 reported an average flux that is twice as large as it should be. This has been corrected in MEM~3.

\subsection{Overall flux}

\begin{figure} \centering
\includegraphics[width=\linewidth]{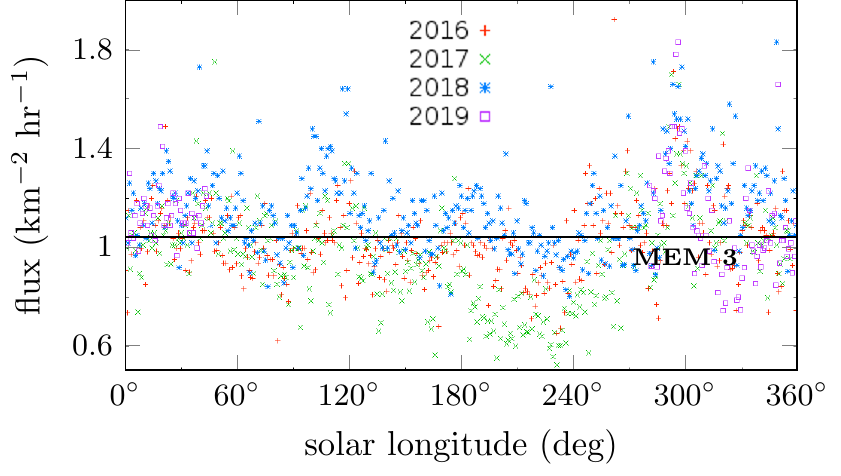}
\caption{Meteoroid flux measured by the CMOR (points) and predicted by MEM~3 (line).}
\label{fig:toaflux}
\end{figure}

The flux predicted by MEM~3 at the top of the Earth's atmosphere is a close match to the flux observed by CMOR. Figure \ref{fig:toaflux} compares MEM~3 to flux measurements taken by CMOR over the past three-plus years at a limiting mass of $1.5 \times 10^{-4}$~g. The overall flux encountered by spacecraft, however, differs between MEMR2 and MEM~3. This is due primarily to the adjustments we have made to meteoroid source strengths and the corrections we have made to gravitational focusing.

The corrected gravitational focusing is a shallower function of altitude, and thus spacecraft in low-Earth orbit experience lower meteoroid fluxes according to MEM~3 than they do according to MEMR2. This reduction is more pronounced for slower meteors; thus, although the flux is lower in LEO, the corresponding speed distribution is faster.

The source strength adjustments we made do not affect the flux at the top of the atmosphere. However, the higher contribution from the helion/antihelion and apex populations, which are less subject to gravitational focusing than the toroidal population, further reduces the effects of gravitational focusing compared to MEMR2.

\begin{figure} \centering
\includegraphics[width=\linewidth]{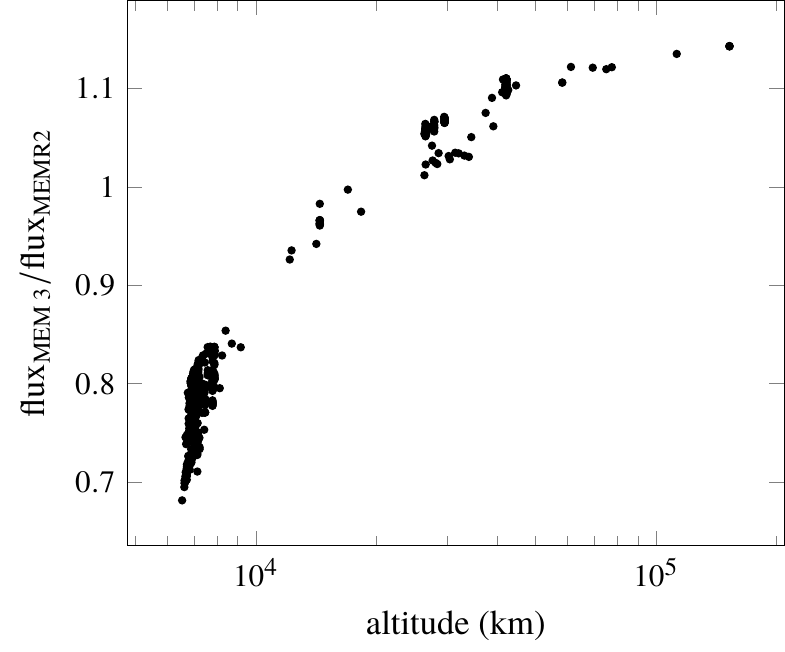}
\caption{Ratio of the flux given by two versions of MEM for randomly chosen spacecraft state vectors.}
\label{fig:ratio23}
\end{figure}

\section{Comparison with \emph{in situ} data}
\label{sec:insitu}

\emph{In situ} measurements of meteoroid impacts that are large enough to be hazardous are rare. Two notable missions that provide impact data in this size range are Pegasus and the Long Duration Exposure Facility (LDEF). \cite{Ehlert2017} compared MEMR2 predictions against large impacts from the LDEF data set and found that it tended to predict fewer craters than were observed, even after an estimate of the orbital debris damage fraction was subtracted. In this section, we use MEM~3 to generate predictions of the number of large craters produced by meteoroid impacts on these two missions and compare our results with the data record.

\subsection{Ballistic limit equations}

MEM reports a mass-limited meteoroid flux; it does not directly predict the number of impact craters on a spacecraft caused by meteoroids. In order to predict the number of craters of a specific limiting size, we must apply ballistic limit equations (or BLEs) to MEM's outputs. BLEs relate the depth and sometimes width of impact craters to the size of the impactor and the properties of both the impactor and target material. They are based on ballistic gun tests, which generally involve metal projectiles and low speeds (by ``low", we mean speeds of a few km~s$^{-1}$ to perhaps 10~km~s$^{-1}$). These ballistic tests are therefore more analogous to orbital debris impacts than they are to meteoroid impacts. However, given that no ballistic gun tests involving meteoroid-like material exist, we employ the same set of BLEs used for orbital debris.

In this work, we apply two sets of BLEs to our simulations. The first is the modified Cour-Palais BLE \citep{Hayashida:1991tm,Christiansen:1992cx}, which offers a relatively simple relationship between the depth of a crater and the impactor's diameter, speed, density, and impact angle. Crater diameter and penetration capability is assumed to be proportional to crater depth (see Sec \ref{sec:ldef} for additional details).  We also apply the Watts \& Atkinson BLEs \cite{1995ldef.symp..523W,1995ldef.symp..287H}, which provide direct relations for the diameter of an impact crater and the thickness of a material that can be penetrated.

\subsubsection{The modified Cour-Palais ballistic limit equation}
\label{sec:CP}

Like \cite{1993Sci...262..550L} and \cite{Ehlert2017}, we use the modified Cour-Palais (hereafter abbreviated as CP) ballistic limit equation \citep{Hayashida:1991tm,Christiansen:1992cx} for a single aluminum sheet to model impact depths. This equation has the form:
\begin{align}
p_c &= 5.24 \, d^{19/18} \, \textrm{BH}^{-1/4} \left( \frac{\rho}{\rho_t} \right)^{1/2}
\left( \frac{v_\bot}{c_t} \right)^{2/3} \label{eq:ble}
\end{align}
where $p_c$ is the depth of the resulting crater in cm in a plate of effectively infinite depth, $d$ is the diameter in cm of the impactor, BH is the Brinell hardness of the target material, $\rho$ and $\rho_t$ are the density of the impactor and the target material, respectively, 
$v_\bot$ is the normal component of the impactor speed relative to the target, and $c_t$ is the speed of sound in the target material. The densities and velocities may be in any units, so long as both densities carry the same units and both speeds carry the same units, but the diameters must be in centimeters. Other units of length would require a change in the constant value of 5.24.

\begin{figure}
\centering
\includegraphics[width=\linewidth]{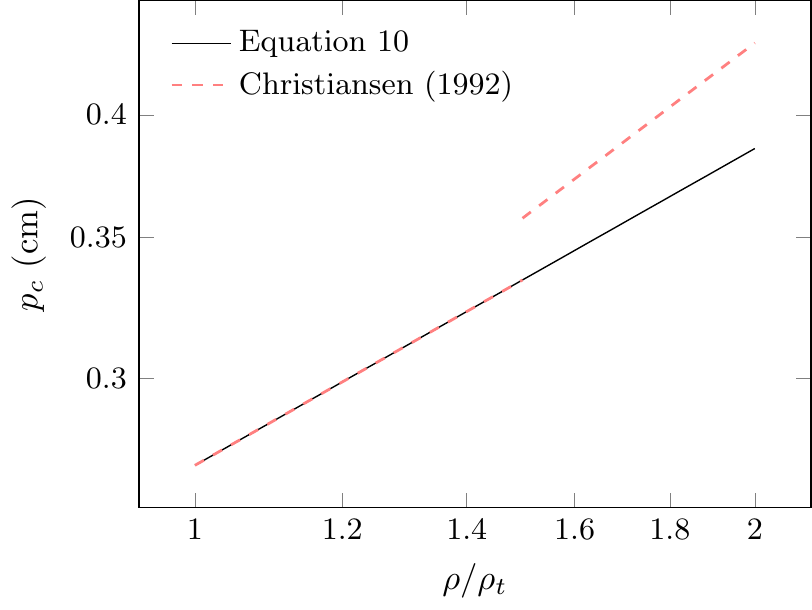}
\caption{Crater depth from a 1-mm meteoroid impacting 6061-T6 aluminum with a normal speed of 15~km~s$^{-1}$.}
\label{fig:disjoint}
\end{figure}

According to \cite{Christiansen:1992cx}, Equation \ref{eq:ble} applies when the impactor's density is less than 1.5 times the target density; for a material density of 2700~kg~m$^{-3}$ for aluminum this limit is 4.05~g~cm$^{-3}$. While meteoroid densities lie below this value on average, some meteoroids exceed this limit. Unfortunately, the high-density extension provided by  \cite{Christiansen:1992cx} appears to be disjoint and therefore probably unphysical (see Figure \ref{fig:disjoint}). We therefore use Equation \ref{eq:ble} for all meteoroid densities.

However, Equation \ref{eq:ble} alone is not sufficient for predicting the number of craters on LDEF nor the number of penetrations of the Pegasus panels. Equation \ref{eq:ble} gives the depth an impactor penetrates into an infinite target; in contrast LDEF crater counts are quoted to a limiting crater \emph{diameter}, and the Pegasus panels have a finite thickness. In each case, we obtain the equivalent limiting crater depth by applying a constant factor to our limiting observable. For LDEF, this factor is the average crater depth-to-diameter ratio, $\xi$, and for Pegasus, this factor is $\tfrac{1}{1.8}$, which is the ratio of penetration depth in an infinite target to the thickness of a finite target that can be penetrated by the same impactor \citep{Hayashida:1991tm}. Thus, for LDEF, $p_c = \xi d_c$, and for Pegasus, $p_c = t/1.8$.

Making our required units of length explicit, we can calculate the particle mass required to produce a crater of depth $p_c$ as follows:
\begin{align}
\frac{d}{\textrm{1 cm}} &= 
	\left[ 
		\frac{p_c}{5.24\textrm{ cm}} 
		h^{1/4}
		\left( \frac{\rho}{\rho_t} \right)^{-1/2}
		\left( \frac{v_\bot}{c_t} \right)^{-2/3}
	\right]^{18/19} 
\end{align}

\subsubsection{The Watts and Atkinson ballistic limit equation(s)}
\label{sec:Watts}

Watts and Atkinson \cite{1995ldef.symp..523W} (hereafter abbreviated as WA) provide a series of equations that describe the depth and width of an impact crater or the thickness of a target that can be penetrated by an impactor. These quantities are not related by a constant multiplicative factor, as the CP BLE assumes, but are instead calculated separately. Furthermore, the WA equations for penetration depth and limiting target thickness are valid only for impacts that exceed a certain minimum velocity:
\begin{align}
    v_0 &= \sqrt{2 Y_t / \rho_t}
           \left(1 + \sqrt{\rho_t / \rho} \right) \\
        &= \sqrt{2 Y_t / \rho}
           \left(1 + \sqrt{\rho / \rho_t} \right) \, ,
\end{align}
where $\rho$ is the impactor density, $\rho_t$ is the target density, and $Y_t$ is the yield strength of the target. We will assume that impacts with speeds less than the minimum fail to produce craters and can be neglected.
Technically, the impact speed must also exceed the critical speed corresponding to the meteoroid ($v_{0,p}$), but the yield strength of a meteoroid is not known. We therefore follow the example of \cite{1995ldef.symp..287H} in assuming that meteoroids have less strength than aluminum and that $v_0$ exceeds $v_{0,p}$.

All WA equations include a ``supralinearity'' correction factor $f$; because this correction brings the WA penetration depth more in line with the empirical CP BLE, we, like \cite{1995ldef.symp..287H}, choose to apply it. For instance, the diameter of an impact crater is given by:
\begin{align}
d_c &= f \cdot d_{0} \, , & \textrm{where} \label{eq:dc} \\
d_{0} &= 1.3235 \, d (c_t/c)^{2/7}
    (v_\bot/v_0)^{4/7}
    \, , & \textrm{and} \label{eq:d0} \\
f &= \left( 1 + \sqrt{2 \Delta/d_{0}} \right)^{-1/3} \, . \label{eq:f} 
\end{align}
The crater diameter thus depends on the impactor diameter, $d$; 
the sound speed within the target material, $c_t$, and 
within the impactor, $c$;
the normal component of the impactor speed relative to the target, $v_\bot = v \cos \theta$;
and a grain size parameter, $\Delta$. The sound speed within a material is given by:
\begin{align}
c &= \sqrt{E/\rho} \, ,
\end{align}
where $E$ is the Young's modulus (also known as the modulus of elasticity). 

Of these quantities, the speed of sound in a meteoroid is the least certain. We first considered a possible value of 2500~m~s$^{-1}$; this is similar to the sound speed in water ice 
and to shear wave speeds in meteorites (1870--2450 m~s$^{-1}$ \cite{2004EM&P...95..361F}). Studies of the effects of porosity or cracks find that porous materials have similar or lower speeds of sound than non-porous materials \cite{2005LPI....36.1856T,2013A&A...554A...4K,2015LPICo1829.6005F}. However, we found that this value produced overly shallow craters (see Section \ref{sec:ldefsims}) and thus adopt the value of 5~km~s$^{-1}$ used by Humes \cite{1995ldef.symp..287H}. 

A formula for the limiting target thickness that can be penetrated by a meteoroid is also provided by \cite{1995ldef.symp..523W}; this is
\begin{align}
t_t &= \frac{f d}{4} \left(
	\frac{1}{6} \frac{\rho}{Y_t} 
	\left(c_{0,t} + \frac{s(v_\bot - v_0)}{1 + \sqrt{\rho_t/\rho}} \right)
	(v_\bot - v_0) \right)^{1/3} \nonumber \\
&+ \frac{f d}{4} \frac{v_\bot}{v_0}
    \sqrt{\frac{Y_t}{\sigma_t}} \label{eq:tt}
\end{align}
where $s_t$ is the stress factor, $\sigma_t$ is the ultimate strength, and $c_{0,t}$ is the sound speed in the unstressed target material (taken by \cite{1995ldef.symp..287H} to be equal to $c_t$).

Inverting the WA damage equations to determine the limiting mass corresponding to a given limiting crater diameter is somewhat less straightforward than it is for the Cour-Palais equation. When the limiting effect is crater diameter, one can combine Equations \ref{eq:dc} and \ref{eq:f} and solve for $d_0$:
\begin{align}
    d_c &= d_0 \left( 1 + \sqrt{2 \Delta/d_{0}} \right)^{-1/3} \, .
\end{align}
This is essentially a septic equation and, as far as the authors have been able to determine, has no analytic solution. However, if $d_c$ is known, this equation can be solved numerically \emph{once} at the beginning of one's analysis. We can then invert Equation \ref{eq:d0} to obtain $d$, and if we once again assume a spherical impactor, we obtain:
\begin{align}
d &= \frac{d_0}{1.3235}
    (c_t/c)^{-2/7} (v_\bot/v_0)^{-4/7} \, . 
\end{align}
Note that in this case, no particular units are assumed.

Inverting Equation \ref{eq:tt} is still more difficult because the supralinearity factor, $f$, is tied to crater diameter, which cannot be determined directly from target thickness. Instead, we define the following unitless quantities:
\begin{align}
x =& \, 1.3235 \, (c_t/c)^{2/7}
     (v_\bot/v_0)^{4/7} \\
y =& \,\left( \frac{\rho (v_\bot - v_0)}{6 Y_t} \left(c_t + 
            \frac{s(v_\bot - v_0)}{1 + \sqrt{\rho_t/\rho}} \right)
	         \right)^{1/3} \nonumber \\
   & \, + \frac{v_\bot}{v_0} \sqrt{\frac{Y_t}{\sigma_t}} \, .
\end{align}
Then, $t_t = y/4 f d$ and $d_0 = x d$, where $x$ and $y$ contain all dependence on variable meteoroid properties such as $v_\bot$ and $\rho$.  Our equation for the limiting target thickness simplifies to:
\begin{align}
t_t &= \tfrac{1}{4} y d \left( 1 + \sqrt{2 \Delta / x d} \right)^{-1/3}
\end{align}
Let us define two additional unitless quantities: $z = y d/4 t_t$ and $b = \sqrt{\Delta y/2 x t_t}$. Then,
\begin{align}
    z^3 &= 1 + b z^{-1/2} \, . \label{eq:z}
\end{align}
Now that we have reduced the inversion of Equation \ref{eq:tt} to solving the above equation, we proceed as follows. 

\begin{figure}
    \centering
    \includegraphics[width=\linewidth]{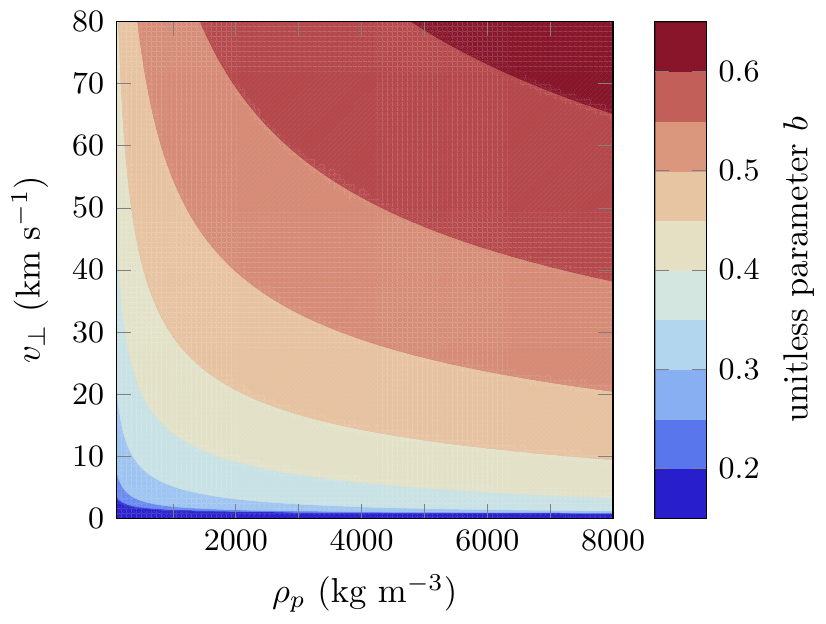}
    \caption{Unitless parameter $b$ as a function of impactor density and normal velocity.}
    \label{fig:pvb}
\end{figure}

First, we determine a range of $b$ values that encompasses the values we encounter in our simulations. We find that $b$ increases monotonically with both $\rho$ and $v_\bot$; see Figure \ref{fig:pvb} for an example calculated using the material properties of 2024-T3 aluminum alloy. Note that when $v_\bot = v_0$, $y = \sqrt{Y_t/\sigma_t}$ and $x = 1.32348 (c_t/c)^{2/7}$; thus, $b$ is constant along this boundary. This minimum value of $b$ is approximately 0.2 for the target properties considered in this paper. We obtain maximum values of $b$ of 0.621 or 0.658, depending on the alloy considered. Thus, the range [0.15, 0.7] should include all values of $b$ corresponding to meteoroids encountered in LEO by Pegasus and LDEF.

\begin{figure}
    \centering
    \includegraphics[width=\linewidth]{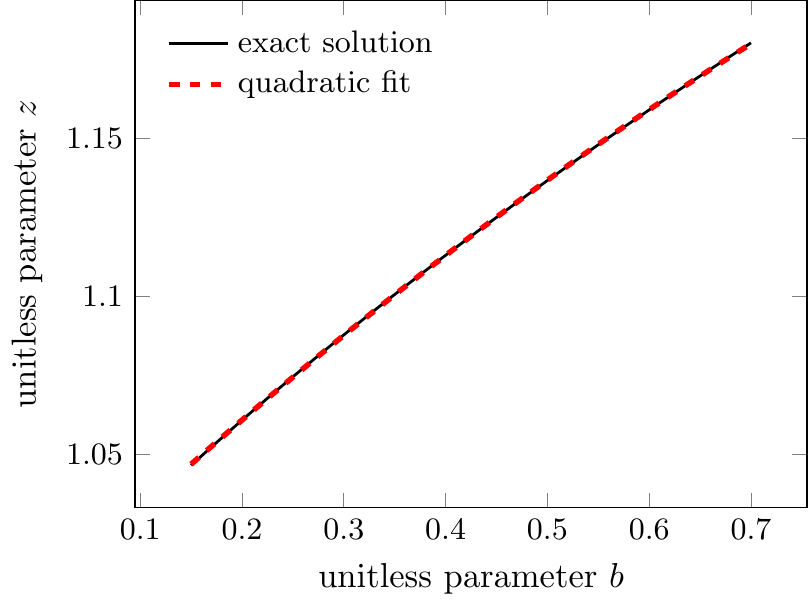}
    \caption{Unitless parameter $z$ as a function of unitless parameter $b$.}
    \label{fig:bz}
\end{figure}

We next solve Equation \ref{eq:z} for all values of $b$ in our range (see Figure \ref{fig:bz}). We find that we can approximate this relationship with a quadratic function:
\begin{align}
    z &\simeq 1.00288 + 0.305355 b - 0.074706 b^2 \, .
    \label{eq:zapprox}
\end{align}
Equation \ref{eq:zapprox} produces results that are within 0.015\% of the true value throughout our range. Thus, once we have determined $b$, we can obtain a close estimate of $z$ using Equation \ref{eq:zapprox}, and convert the result to a limiting particle diameter as follows:
\begin{align}
    d &= 4 z t_t / y
\end{align}

\subsubsection{Weighting to a constant crater diameter}
\label{sec:weighting}

We will assume spherical meteoroids throughout our analysis; the limiting mass is therefore
\begin{align}
    m &= \pi \rho d^3 / 6 \, . \label{eq:mp}
\end{align}
Once we have determined the limiting mass, we can then weight the results of our MEM runs to a constant limiting crater diameter. For each mass-limited flux component $F_m$, we calculate the corresponding crater-limited flux component as follows:
\begin{align}
F_c &= F_m \frac{F_G(m)}{F_G(10^{-6}\textrm{ g})} \, .
\end{align}
The quantities $m$ and $F_G(m)$ must be calculated for every radiant-velocity bin in our flux file and for every density bin in our density distribution file. Both the total flux incident on a particular surface and the limiting mass depend on the impact angle. Our final predicted crater count will be 
\begin{align}
N_{c,i} &= A_i \Delta t \sum_{j,k,l} F_{m,j,k,l} \cos{\alpha_{i,j}} \frac{F_G \circ m(v_k, \alpha_{i,j}, \rho_l)}{F_G(10^{-6}\textrm{ g})}
\end{align}
where $\circ$ indicates function composition, $A_i$ is the area of surface $i$, $\Delta t$ is the exposure time, $\alpha_{i,j}$ is the angle between normal vector $i$ and meteoroid radiant $j$, and $F_{m,j,k,l}$ is the mass-limited flux in angular bin $j$, velocity bin $k$, and density bin $l$. Alternatively, we may omit $A_i$ and $\Delta t$ and instead compute the cratering or penetration rate.

\subsection{Pegasus}
\label{sec:pegasus}

The three Pegasus satellites, launched in 1965, used large arrays of penetration detectors to measure the meteoroid impact flux. These detectors covered both sides of two extendable wings on each spacecraft and consisted of a surface layer of aluminum alloy on top of layers of dielectric mylar and vapor-deposited copper maintained at a constant voltage \cite{Clifton:1966uw}. Meteoroids penetrating to the dielectric/copper layers were detected in the form of a discharge. Thus, the Pegasus detectors were penetration-limited; impacts that only partially penetrated the aluminum did not cause discharges and were not detected. Each capacitor stopped working when punctured, and the usable area decreased over time. 

No meteoroid properties were measured other than the ability to penetrate these detectors: velocity, density, and impact angle were all unknown. On the other hand, the sheer size of these detector arrays (over 600 square meters), combined with their 2 year operation, produced one of the best measurements of the near-Earth meteoroid impact flux ever made.  The Pegasus data has the additional advantage of being uncontaminated by orbital debris.  We assume that orbital debris was a negligible source of impacts in 1965, and therefore assume that all impacting particles are meteoroids.

\begin{table*} \centering
\begin{tabular}{cclll}
&& Pegasus & LDEF & LDEF (LB93) \\
\hline
target density & $\rho_t$ &
    2.78 g~cm$^{-3}$ & 
    2.7 g cm$^{-3}$ & 2.7 g cm$^{-3}$ \\
Brinnel hardness & BH & 
    120 & 95 & 90 \\
Young's modulus & $E_t$ & 
    73.1 GPa & 68.9 GPa & -- -- \\ 
sound speed & $c_t$ &
    5.13 km s$^{-1}$ & 
    5.05 km s$^{-1}$ & 6.1 km s$^{-1}$ \\
yield strength & $Y_t$ &
    345 MPa & 276 MPa & -- -- \\
ultimate strength & $\sigma_t$ &
    483 MPa & 310 MPa & -- -- \\
stress factor & $s$ &
    1.6 & 1.42 & -- -- \\
grain size factor & $\Delta$ &
    50~$\mu$m & 50~$\mu$m & -- -- \\
depth-to-diameter ratio & $\xi$ & -- -- & 0.5 & 0.527
\end{tabular} \vspace{0.2cm}
\caption{Material properties assumed for Pegasus, LDEF, and those used by \protect\citep{1993Sci...262..550L} and \protect\cite{Ehlert2017}.}
\label{tab:6061}
\end{table*}

\subsubsection{Data}

Each Pegasus spacecraft had 208 detectors that took up 194.5~m$^2$ of surface area \cite{Clifton:1966uw}. The detectors had three thicknesses -- 0.4~mm, 0.2~mm, and 0.038~mm -- with larger surface areas dedicated to the thicker detectors \cite{Clifton:1966uw}. A typical thickness for a spacecraft's external surface is about 1~mm; thus, the Pegasus missions probed the meteoroid environment at sizes a little smaller than would typically endanger a spacecraft. Nevertheless, we will test MEM against the penetration rate experienced by the 0.4~mm detectors. We will not simulate penetration of the thinner detectors, as they correspond to even smaller particles. The material properties for the aluminum plates are given in Table \ref{tab:6061}.

Orientation information for the three spacecraft does not appear to have been preserved. However, the angle of the Sun with respect to the sensor plane of Pegasus I and the rotational axis of Pegasus II in June of 1965 is shown in \cite{Naumann65}; both angles vary by at least 90$^\circ$. Based on these angles, \cite{Naumann65} concluded that the entire celestial sphere was swept by Pegasus II's sensor axis each day. We will therefore treat the Pegasus satellites as randomly tumbling in our analysis.

Clifton and Naumann \cite{Clifton:1966uw} reported a penetration flux of 0.00487~m$^{-2}$~day$^{-1}$ for the 0.4-mm detectors. This was based on the 201 penetrations recorded by Pegasus II and III before the end of 1965. Pegasus I exhibited anomalous behavior and was not analyzed by \cite{Clifton:1966uw}. (An earlier report by Naumann \cite{Naumann65} gave a rate of 0.0021~m$^{-2}$~day$^{-1}$ for Pegasus I and 0.0035~m$^{-2}$~day$^{-1}$ for Pegasus II based on 4 and 30 detections, respectively, between February and July, 1965.) Although the mission continued until 29 August 1968, there were no subsequent analyses of the meteoroid flux. 

\begin{figure*}
    \centering
    \includegraphics[width=\linewidth]{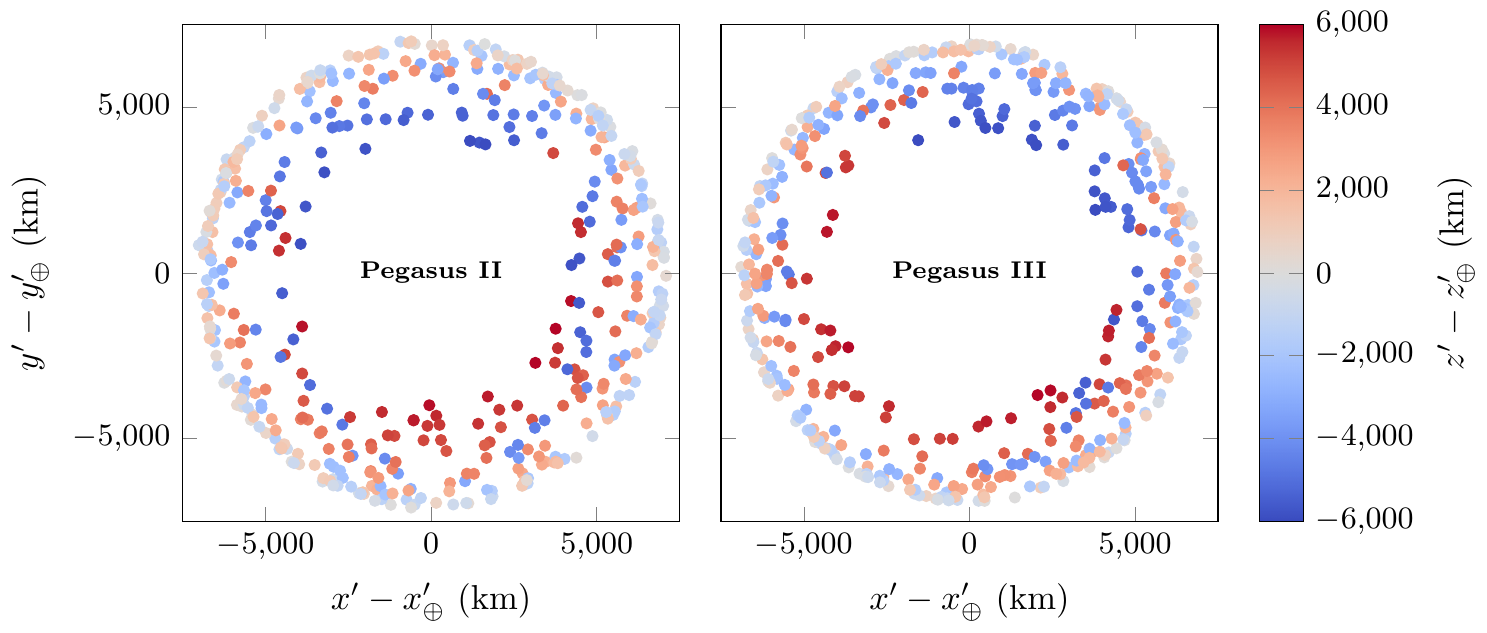}
    \caption{Position of Pegasus II and III for each randomly selected state vector used in this analysis.}
    \label{fig:pegran}
\end{figure*}

\subsubsection{Simulations}

Because Pegasus~I is not included in the flux quoted by \cite{Clifton:1966uw}, we also exclude it from our analysis. We generated trajectories for the two remaining satellites by converting two-line element (TLE) sets to state vectors at random times between the epoch of the first available TLE and Jan 1, 1966. Times were selected randomly to avoid aliasing between the sampling interval and the orbital period. The resulting state vectors evenly cover each orbit; Figure \ref{fig:pegran} plots the position of each spacecraft relative to the Earth in the corotating frame. The trajectory was limited to 1965 to better mimic the data used by \cite{Clifton:1966uw}. The two satellites have similar orbits: Pegasus III has a slightly shorter orbital period than Pegasus II, but Pegasus II has a slightly larger orbital eccentricity and reaches a lower altitude at perigee.

We ran MEM~3 with these state vectors in high-fidelity mode, with a 3$^\circ$ angular resolution and a 1~km~s$^{-1}$ velocity resolution. We opted to output an average igloo file; this is an output file format with a variable number of azimuthal bins per row in elevation (see, e.g., \cite{Krisko2010}). The smaller number of bins in the threat igloo reduced the time required for our subsequent calculations.

As mentioned previously, we treat the Pegasus satellites as randomly tumbling. Thus, we construct a series of random orientation vectors; for each orientation vector, we also draw 10 random densities for each population using Equation \ref{eq:logrho}. We then calculate the angle between the random surface normal vector and each angular mid-point reported in the igloo files, and calculate the penetration-limited flux as outlined in Sections \ref{sec:Watts} and \ref{sec:weighting}. We repeat this process for additional random orientation vectors and sets of random densities until the results converge (about 500 iterations are required to obtain convergence to the third decimal place).

\begin{figure}
    \centering
    \includegraphics[width=\linewidth]{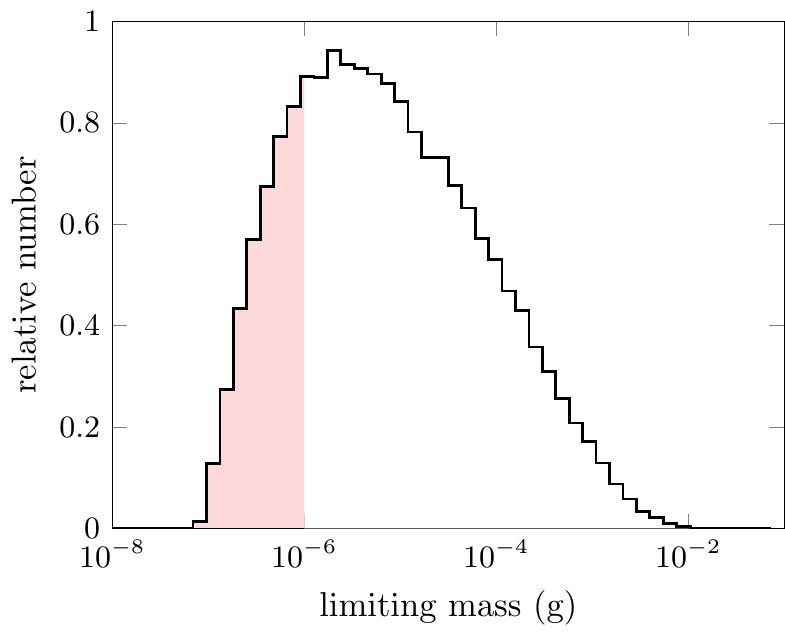}
    \caption{Histogram of limiting masses for one Pegasus~II state vector.}
    \label{fig:mhist}
\end{figure}

Because Pegasus probes particle sizes below MEM~3's cutoff of 10$^{-6}$~g, we estimate the effect of incompleteness by performing our simulations twice. In the first case, we enforce a minimum limiting mass, where any calculated limiting masses falling below this limit are replaced by $10^{-6}$~g. In the second case, we use the Gr\"{u}n flux curve to extrapolate to masses well below the MEM~3 mass limit. Figure \ref{fig:mhist} plots a rough distribution of limiting masses over all modeled velocities and densities for Pegasus II; a sizeable fraction of these masses fall below our model's limit of $10^{-6}$~g.

\subsubsection{Results}

\begin{figure}
    \centering
    \includegraphics[width=\linewidth]{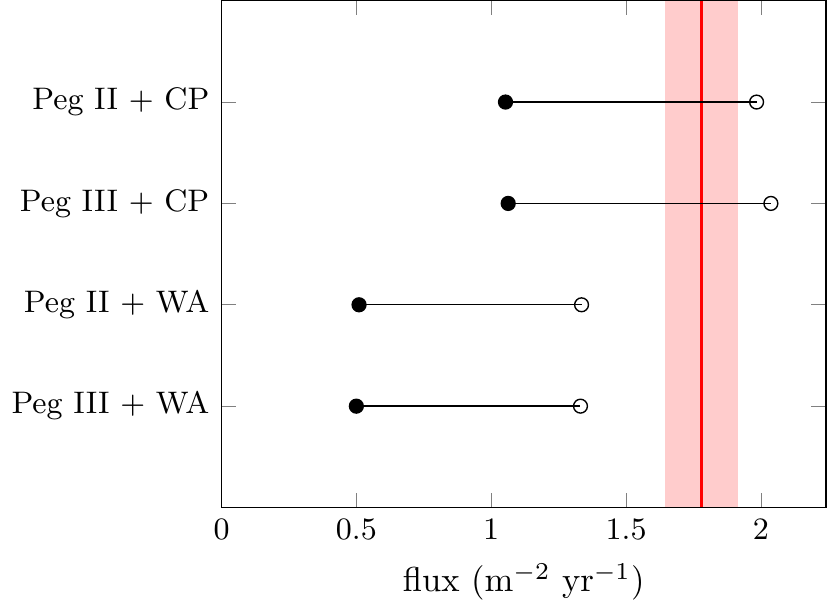}
    \caption{Penetration flux predicted by MEM~3 (intervals) for the Pegasus II and III spacecraft (vertical line).}
    \label{fig:peg}
\end{figure}

Figure \ref{fig:peg} summarizes our predicted penetration rates for the Pegasus spacecraft. Four sets of calculations are represented in this plot; we calculate rates for both spacecraft and our two sets of BLEs. For each spacecraft-BLE combination, we handle the limiting mass in one of two ways. First, we restrict our limiting mass to $10^{-6}$~g or larger, regardless of whether the detection panels can be penetrated by smaller particles. This gives us a lower estimate of the penetration rate and these values appear in Figure \ref{fig:peg} as solid black points. Second, we include limiting masses smaller than $10^{-6}$~g and extrapolate our flux results to those masses using the Gr\"{u}n et al.~flux curve. The resulting values are shown in Figure \ref{fig:peg} as open circles. The flux measured by \cite{Clifton:1966uw} is shown as a vertical red line, and their 1-$\sigma$ uncertainties are represented by the shaded region.

Our nominal predictions for Pegasus were low compared to observations, although our extrapolated results (which included an estimate of the effects of meteoroids below our mass limit) exceeded the number of penetrations recorded by Pegasus when using the CP BLE. Use of the WA BLE led to underpredicted penetrations for all cases. Pegasus II and III showed no significant differences in the predicted penetration rate.

If particles smaller than $10^{-6}$~g have higher speeds or densities than those modeled by MEM, it is possible for the penetration rate to exceed those calculated here. Separately, it is possible that the penetration data from Pegasus are contaminated by false positives; \cite{Clifton:1966uw} noted that it was possible that electrical shorts due to causes other than particle penetrations were present in the data, although they deemed it unlikely.

\subsection{LDEF}
\label{sec:ldef}

The Long Duration Exposure Facility (LDEF) was designed to collect a variety of space environment data; the spacecraft orbited the Earth for almost 6~years before being retrieved for analysis. LDEF followed a nearly circular orbit with an altitude of 500~km and inclination of 28.5$^\circ$ \cite{1990LPI....21.1385Z} and maintained a fixed orientation relative to its orbit (i.e., the same surface faced the Earth's surface for the entire mission). The largest meteoroid impact experiment on LDEF had a surface area of 9.34~m$^2$; this large surface size, when combined with the long mission duration, make LDEF one of the few \emph{in situ} experiments to measure meteoroid impacts in the threat regime.

Its experiments included the Meteoroid and Space Debris Impact Experiment, which exposed a set of aluminum plates to the space environment \cite{1991NASCP3134..399H}, and the Chemistry of Micrometeoroids Experiment (CME) \cite{Horz95}, which exposed both aluminum and gold targets. Later, meteoroid crater data were also obtained from frame components (called the ``intercostals'' and the largest source of data outside of dedicated experiment plates) \cite{ZOLENSKY1991} and other surfaces such as clamps.

By the 90s, orbital debris was present and contributed to the crater count on LDEF. As a result, the CME data, which could, to some degree, distinguish between natural and man-made impactors, will be important to our analysis.

\subsubsection{Data}

\begin{figure}
\centering
\includegraphics[width=\linewidth]{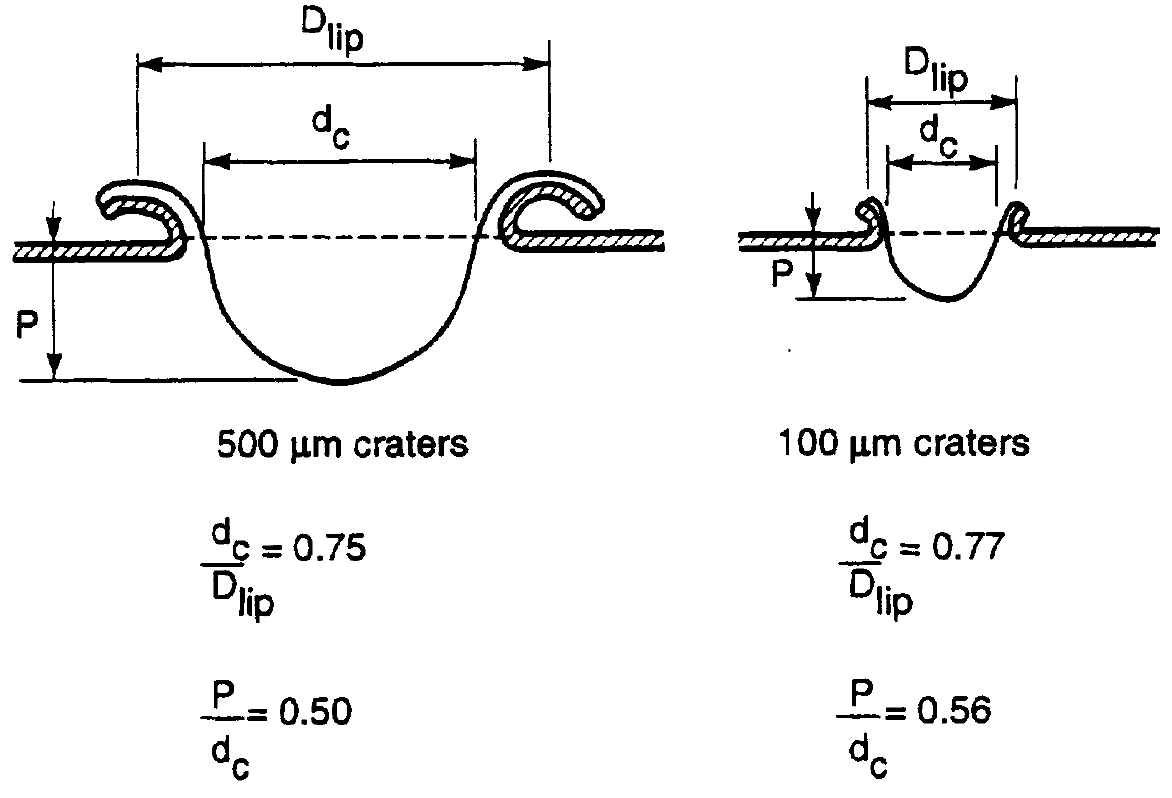}
\caption{Typical morphology of craters on LDEF; diagram from \cite{1995ldef.symp..287H}.}
\label{fig:crater}
\end{figure}

Humes \cite{1995ldef.symp..287H} provides crater counts for the Meteoroid and Space Debris Impact Experiment; in total, this experiment produced 4341 craters that are at least 100~$\mu$m, or 0.1~mm, in diameter. This includes the ``lip'' of the crater (see Figure~\ref{fig:crater}); the internal diameter is smaller, and the depth smaller still. According to \cite{1995ldef.symp..287H}, the depth of a 100~$\mu$m lip-diameter crater is only 43~$\mu$m (see Figure~\ref{fig:crater}): too small to pose a threat to spacecraft, and too small to be modeled by MEM. We therefore follow the example of \cite{Ehlert2017} and make use only of the 1000~$\mu$m data.

Figure \ref{fig:crater} summarizes the crater features discussed by \cite{1995ldef.symp..287H}: namely, that large craters have a roughly hemispherical interior, with the depth equaling half the interior diameter; that the interior diameter is about three-quarters the ``lip'' diameter; and that small craters are proportionally deeper and have a slightly smaller lip. 

This depth-to-diameter ratio of 0.5 differs from the value of 0.527 used by \cite{1993Sci...262..550L} and \cite{Ehlert2017}. The earlier paper \cite{1993Sci...262..550L} claims that the 0.527 value was based on a 1990 preliminary report by See et al.~\cite{See1990Report} and supposedly did not vary with crater size. As far as we can determine, other articles that cite a depth-to-diameter ratio deeper than 0.5 \cite[i.e.,][]{1995LPI....26..109B} do so based on LDEF's 100~$\mu$m crater data.  We will assume a constant depth-to-diameter ratio of 0.5 in this work when using the CP BLEs; the WA BLEs allow us to compute crater diameter directly and thus allow for variable depth-to-diameter ratios naturally.

Humes also claims, based on a ballistic model by Watts and Atkinson \cite{1995ldef.symp..523W}, that a meteoroid density of 2500~kg~m$^{-3}$ is required to produce the depth-to-diameter ratios seen for the LDEF craters, and that the 500~kg~m$^{-3}$ density quoted by \cite{SP8013} produce craters that are too shallow.  MEM~3 contains two density populations, one of which is clustered around $\sim 3800$~kg~m$^{-3}$ and the other around $\sim 860$~kg~m$^{-3}$, and thus spans both values.

Due to the presence of orbital debris in the data, we use CME data to estimate the fraction of impacts that are due to meteoroids. CME consisted of seven panels of high-purity gold on LDEF's wake-facing surface (side 3) and 6 panels of aluminum (alloy Al 1100) on one of the forward-facing surfaces (side 11). The purpose of the experiment was to collect impactor residues from the craters left on these surfaces and extract compositional information using a scanning electron microscope and x-ray spectral analysis \cite{Horz95}. The presence of Fe-Ni-Cr, Zn-Ti-Cl, Ag, Cu, Pb-Sn, and Al were interpreted as signatures of orbital debris, while Si, Mg, Fe, Fe-Ni-sulfide melts, and well-mixed, fine-grained matrices, were assumed to be from natural particles.

Because gold is not a significant component of either orbital debris or meteoroids, residues from the wake-facing surface could be classified as either man-made or natural with a greater degree of confidence than residues from the ram-facing surface. The aluminum panels, besides being compositionally indistinguishable from aluminum orbital debris, also suffered varying levels of contamination. For instance, every crater on aluminum plate E00H contained Si, Mg, Fe, and other contaminants, leading it to be excluded entirely from the H\"{o}rz et al.~analysis \cite{Horz95}. CME data are available for two sides of LDEF: side 3, which faces the wake direction, and side 11, which is on the ram-facing half of the spacecraft. Based on these CME results, \cite{Ehlert2017} assumed that 10\% of impacts on side 3 and 45\% of impacts on side 11 were due to orbital debris. All impacts on the zenith-facing surface are attributed to meteoroids.

While craters were also collected from the intercostals, we, like \cite{Ehlert2017}, have opted to exclude these data. This is for several reasons: the surface area is smaller, no data exists for zenith- or nadir-facing surfaces, and the limiting crater diameter (640 microns) is smaller. We therefore model only the largest craters from sides 3, 11, and 13 of the Meteoroid and Space Debris Impact Experiment; the material properties of the aluminum panels is given in Table \ref{tab:6061}.

We found some disagreement in the material properties used by \cite{1993Sci...262..550L} and those cited by \cite{1995ldef.symp..287H}. The latter quotes values that are in closer agreement with those listed in current material data sheets. Table \ref{tab:6061} compares the two sets of values; those used by \cite{1993Sci...262..550L} are marked ``LB93.'' Note that we assume that the impact crater is hemispherical; i.e., we assume that $p_c = \tfrac{1}{2} d_c$, where $d_c$ is the diameter of the crater.

\subsubsection{Simulations}
\label{sec:ldefsims}

A series of TLEs describing LDEF's orbit (the same file used by \cite{Ehlert2017}) was converted to state vectors with a sampling interval of 910 minutes, which corresponds to 9.68 orbits. This trajectory file covers the entirety of LDEF's operational lifetime.

We ran MEM~3 in high-fidelity mode, with a 3$^\circ$ angular resolution and a 1~km~s$^{-1}$ velocity resolution. We used a body-fixed output coordinate frame that matches LDEF's orientation relative to its orbit. We again used the threat igloo output file to compute the flux on various surfaces.

\begin{figure}
\centering
\includegraphics[width=\linewidth]{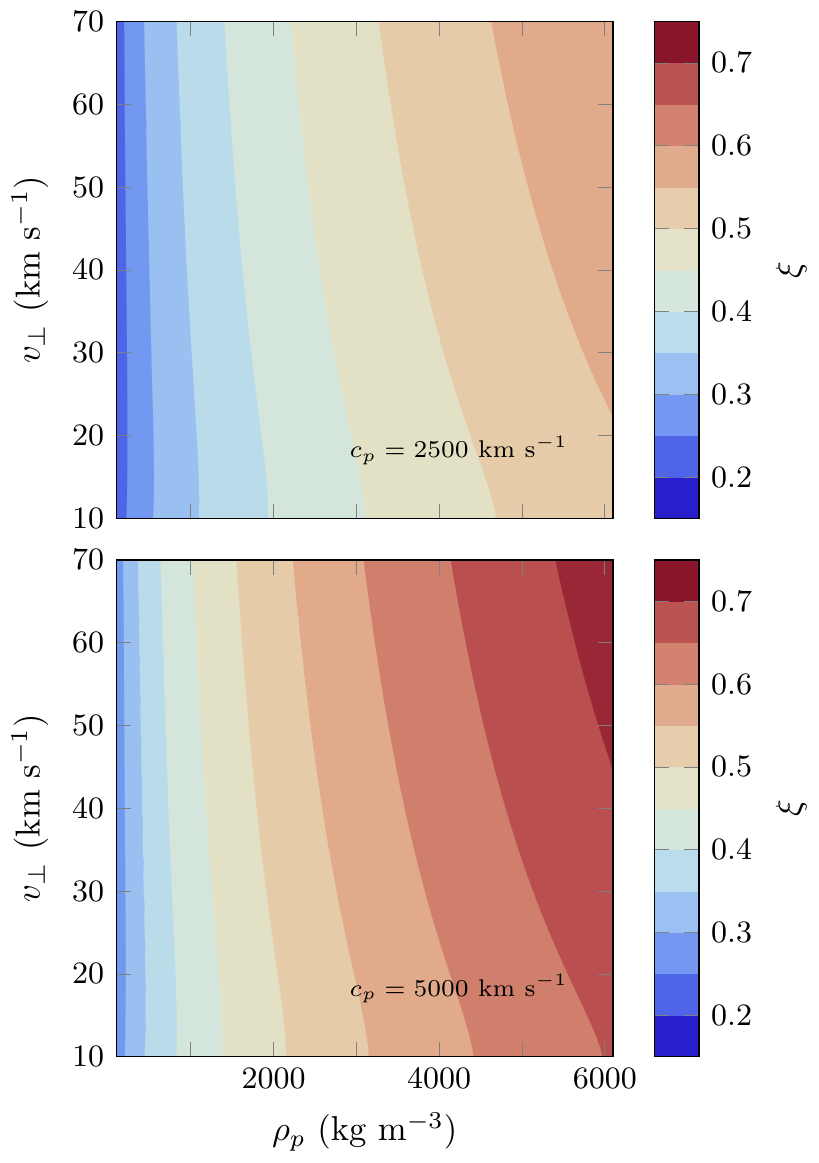}
\caption{Depth-to-diameter ratio, $\xi$, as a function of impactor density and normal velocity.}
\label{fig:cp}
\end{figure}

Reference \cite{1995ldef.symp..287H} claims that high meteoroid densities (i.e., 2500~kg~m$^{-3}$ rather than 500~kg~m$^{-3}$) are required to produce the observed depth-to-diameter ratio of 0.5. In fact, both particle density and particle velocity influence the depth-to-diameter ratio, although particle density does appear to have a much stronger effect (see Figure \ref{fig:cp}). For a meteoroid sound speed of 2.5~km~s$^{-1}$, we attain depth-to-diameter ratios of 0.5 only for the densest, fastest meteoroids. If we instead assume a meteoroid sound speed of 5~km~s$^{-1}$, then our denser population of meteoroids (i.e., the helion and antihelion sources) will produce craters a little deeper than $\xi = 0.5$, while the apex and toroidal sources will produce craters with values of $\xi$ closer to 0.4. This motivated us to adopt a meteoroid sound speed of 5~km~s$^{-1}$ throughout our analysis, despite the fact that this speed is more characteristic of metals than natural materials.

\subsubsection{Results}

\begin{figure}
    \centering
    \includegraphics[width=\linewidth]{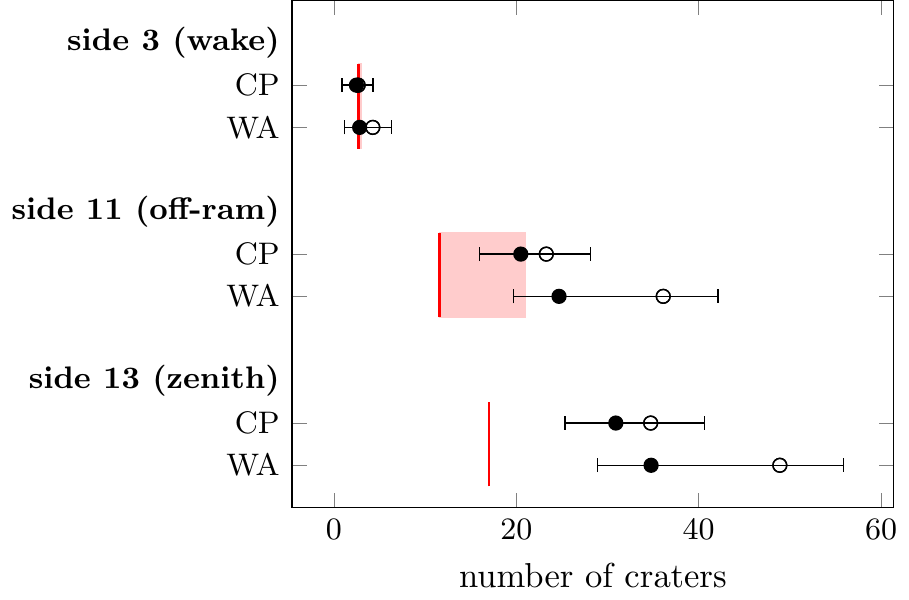}
    \caption{Number of craters predicted by MEM~3 (points) for three sides of the LDEF spacecraft (vertical lines).}
    \label{fig:ldef}
\end{figure}

Figure \ref{fig:ldef} summarizes our predicted penetration rates for the LDEF spacecraft. We have calculated the number of craters on the three sides of the spacecraft (sides 3, 11, and 13) for which we have orbital debris fraction estimates using our two sets of BLEs (CP and WA). As with our Pegasus analysis, we first restrict our limiting mass to $10^{-6}$~g or larger to calculate a nominal prediction (solid black circle); we then also extrapolate to smaller masses (open circle). The sampling error ($\sqrt{n}$) extends our uncertainty and is shown by the error bars. The vertical red lines in Figure \ref{fig:ldef} mark the number of craters that we estimate are due to meteoroids. We also include a shaded interval that spans the range from this estimate to the total number of craters due to any impactor.

Our predictions exceed the number of craters on LDEF by about a factor of two on two of the three sides. Side 11 is estimated to be significantly contaminated by orbital debris; if this contamination level is lower than we have estimated, the data could be in better agreement with our predictions. However, on side 13, the orbital debris contamination should be minimal and thus our predictions are an unambiguous overestimate.  The greater disagreement on the zenith-facing surface than on the other surfaces could be due to a difference in the assumed speed distribution \cite{1991NASCP3134..569Z}; this will be the focus of a future study.

We also find that including masses smaller than MEM's threshold of $10^{-6}$~g makes a more significant difference when using the WA BLE than when using the CP BLE. This is likely because, according to the WA BLE, slow, low-density particles create wide craters and thus smaller particles may more readily contribute to a diameter-limited crater count.

\subsection{Discussion}

In this section, we compared predictions from MEM~3 with two \emph{in situ} experiments: the Pegasus satellites and the Long Duration Exposure Facility (LDEF). The penetration rate MEM~3 predicts for Pegasus is a factor of 2-3 too low, while the number of craters MEM~3 predicts for LDEF is about a factor of 2 too high. This result depends on the BLE used: use of the CP BLE produced more predicted penetrations for Pegasus than the WA BLE, but fewer craters on LDEF than the WA BLE. The results were also sensitive to the method for handling particles smaller than MEM's mass limit of $10^{-6}$~g; extrapolating to smaller particles always results in more predicted penetrations and craters. 

The uncertainty in the overall meteoroid flux at 1~au is sometimes cited as being a factor of approximately 3 \cite{Drolshagen:2015uk}. The results presented in this paper certainly seem to indicate that this is a reasonable estimate; given the lack of information on meteoroid physical properties, such as sound speed, the uncertainties could conceivably be larger. 

Further comparisons between environment models and existing \emph{in situ} data are needed. Possible additional data sources are Space Shuttle hypervelocity impacts \cite{Stucky:2019ud}, craters on returned ISS hardware \cite{Hyde:2017tf}, the European Retrievable Carrier (EURECA) \cite{1995AdSpR..16...85D}, or Hubble solar panels \cite{McDonnell2005}. These data involve non-aluminum targets and/or spacecraft shapes with concavities, and thus their analysis will be more complex than that presented here.

\section{Conclusions}

MEM~3 is the latest version of NASA's Meteoroid Engineering Model and introduces a number of improvements over previous versions. It includes: new density distributions based on recent research, a corrected gravitational focusing and planetary shielding algorithm, properly preserved correlations between meteoroid speed and direction, and adjustments of the relative strength of the constituent meteoroid orbit populations that better match meteor radar data. The code has a reduced run time and expanded error-trapping abilities, and submodels have been eliminated in favor of a single, streamlined interface. MEM~3 can model the meteoroid environment near Mercury, Venus, and Mars, and a command-line version of the code is available for Windows, Linux, and Mac operating systems.

The meteoroid flux predicted by MEM~3 at the top of the Earth's atmosphere is a close match to that observed by CMOR. Due to the source strength adjustments made in this version, it also more closely matches the observed directionality of the meteoroid environment than previous versions of MEM. However, it was not possible to match the observed speed distribution \cite{2017P&SS..143..209M,2018M&PS...53.1292M} with the current orbital populations; a re-investigation of these orbital populations will be the subject of future work.

MEM~3 predicts a level of damage that is in rough agreement with the Pegasus and LDEF \emph{in situ} data. MEM~3 underpredicts the number of penetrations detected by Pegasus and overpredicts the number of craters collected from LDEF. This disagreement is approximately a factor of 2 in both cases, consistent with an overall uncertainty in the environment of a factor of 3. Some of this disagreement could be associated with the interpretation of the \emph{in situ} data; it is possible that the Pegasus data contain false positives due to non-impact-induced shorts, and it is also possible that orbital debris plays a smaller role than estimated at the large crater sizes considered here.

Further comparisons between environment models and existing \emph{in situ} data are needed, but this uncertainty will likely only be resolved by the collection of additional \emph{in situ} data specifically in the threat regime. 

\section*{Acknowledgement}

This work was supported in part by NASA contract 80MSFC18C0011.

AVM would like to thank their co-authors, Rob Suggs, and Kirk Heuiser for beta-testing MEM~3. 

\bibliography{local}

\end{document}